\documentclass[fleqn,usenatbib]{mnras}

\usepackage{newtxtext,newtxmath}

\usepackage[T1]{fontenc}
\usepackage{ae,aecompl}

\usepackage{graphicx}	
\usepackage{amsmath}	
\usepackage{amssymb}	
\usepackage{bm}  
\usepackage{subfig}
\usepackage{float}

\usepackage{enumitem}
\usepackage[pdftex,dvipsnames]{xcolor}
\usepackage[linecolor=MidnightBlue,backgroundcolor=MidnightBlue!25,bordercolor=MidnightBlue,prependcaption]{todonotes}

\usepackage{braket}

\usepackage{emptypage}

\usepackage{lipsum}

\newcommand{\eqq}[1]{Equation~(\ref{#1})}

\newcommand{\ie}{{\it i.e.}}
\newcommand{\eg}{{\it e.g.}}

\newcommand{\likeli}{\ensuremath{\mathcal{L}}}
\newcommand{\photoz}{photo-$z$}
\newcommand{\features}{\ensuremath{\mathbfit{F}}}
\newcommand{\redshifts}{\ensuremath{\mathbfit{z}}}
\newcommand{\types}{\ensuremath{\mathbfit{t}}} 
\newcommand{\positions}{\ensuremath{\bm{\mathit{\theta}}}}
\newcommand{\fields}{\ensuremath{\bm{\mathit{\delta}}}}
\newcommand{\biases}{\ensuremath{\mathbfit{b}}}
\newcommand{\fractions}{\ensuremath{\mathbfit{f}}}
\newcommand{\counts}{\ensuremath{\mathbfit{N}}}
\newcommand{\priorcounts}{\ensuremath{\mathbfit{M}}}
\newcommand{\nzt}{\ensuremath{{N}_{zt}}}
\newcommand{\mzt}{\ensuremath{{M}_{zt}}}
\newcommand{\des}{\textit{DES}}

\definecolor{myorange}{RGB}{255, 104, 51}

\title[Redshifts with colors and clustering]{Redshift inference from the combination of galaxy colors and clustering in a hierarchical Bayesian model -- Application to realistic $N$-body simulations}

\author[Alarcon, S\'anchez, Bernstein \& Gazta\~naga]{
Alex Alarcon$^{1,2}$\thanks{Corresponding author: alexalarcongonzalez@gmail.com},
Carles S\'anchez$^{3}$\thanks{Corresponding author: carless@physics.upenn.edu},
Gary M. Bernstein$^{3}$
and Enrique Gazta\~naga$^{1,2}$
\vspace{2mm}
\\
$^{1}$ Institut d'Estudis Espacials de Catalunya (IEEC), 08193, Barcelona, Spain\\
$^{2}$ Institute of Space Sciences (ICE, CSIC), Campus UAB, Carrer de Can Magrans, s/n, 08193, Barcelona, Spain\\
$^{3}$ Department of Physics \& Astronomy, University of Pennsylvania, 209 S. 33rd St., Philadelphia,
PA 19104, USA
}

\vspace{1mm}

\date{\today}

\pubyear{2019}

\begin{document}
\label{firstpage}
\pagerange{\pageref{firstpage}--\pageref{lastpage}}
\maketitle

\begin{abstract}
Photometric galaxy surveys constitute a powerful cosmological probe but 
 rely on the accurate characterization of their redshift distributions using only broadband imaging, and can be very sensitive to incomplete or biased priors used for redshift calibration.
 \citet{Sanchez2019} presented a hierarchical Bayesian model which estimates those from the robust combination of prior information, photometry of single galaxies and the information contained in the galaxy clustering against a well-characterized tracer population. In this work, we extend the method so that it can be applied to real data, developing some necessary new extensions to it, especially in the treatment of galaxy clustering information, and we test it on realistic simulations. After marginalizing over the mapping between the clustering estimator and the actual density distribution of the sample galaxies, and using prior information from a small patch of the survey,  we find the incorporation of clustering information with \photoz's to tighten the redshift posteriors, and to overcome biases in the prior that mimic those happening in spectroscopic samples. The method presented here uses all the information at hand to reduce prior biases and incompleteness. Even in cases where we artificially bias the spectroscopic sample to induce a shift in mean redshift of $\Delta \bar z \approx 0.05,$ the final biases in the posterior are $\Delta \bar z \lesssim0.003.$
This robustness to flaws in the redshift prior or training samples would constitute a milestone for the control of redshift systematic uncertainties in future weak lensing analyses. 
	
\end{abstract}




\section{Introduction}
\label{sec:intro}

Galaxy surveys provide key information about the large-scale structure of the Universe, constituting one of the most powerful probes for testing cosmological models. There exist two main categories of surveys. On one hand, spectroscopic surveys such as 
\textit{2dF} \citep{Colless2001}, 
the \textit{VIMOS-VLT Deep Survey} \citep{LeFevre2005}, 
\textit{WiggleZ} \citep{Drinkwater2010}, 
\textit{Baryon Oscillation Spectroscopic Survey} \citep{Dawson2013} or 
\textit{Dark Energy Spectroscopic Instrument} \citep{DESI2016} provide three-dimensional information about the galaxies they measure, but they are expensive in time and resources. On the other hand, imaging or photometric surveys like 
the \textit{Sloan Digital Sky Survey} \citep{York:2000gk}, 
\textit{PanSTARRS} \citep{Kaiser2000}, 
the \textit{Kilo-Degree Survey} \citep[\textit{KiDS},][]{Jong2013}, 
the \textit{Dark Energy Survey} \citep[\textit{DES},][]{Flaugher2015}, 
the \textit{Hyper-Suprime-Cam} survey \citep[\textit{HSC},][]{HSC2012}, 
or the \textit{Large Synoptic Survey Telescope} \citep[\textit{LSST},][]{LSST2012}
use less time per galaxy, and enable weak gravitational lensing measurements via galaxy shapes --- but provide only a crude view of the line-of-sight dimension of the Universe, since galaxy redshifts are estimated from only their observed broadband spectra. 

In order to perform unbiased cosmological analyses of imaging surveys it is critical to characterize the redshift distributions $n(z)=\mathrm{d}N/\mathrm{d}z\,\mathrm{d}A$ of the corresponding galaxy samples, and unaccounted errors in such characterization will directly lead to biases in the cosmological parameter estimation \citep{Huterer2006,Hildebrandt2012,Cunha2012,Benjamin2013,Huterer2013,Bonnett2015,Hoyle2017,Hildebrandt2017}. Recently, there has been a number of comparisons between cosmological parameters obtained from imaging suveys \citep{Hildebrandt2018,Troxel2018,Hikage2018} and the cosmic microwave background \citep{PlanckCollaboration2018} which have claimed discrepancies of up to $3.2\sigma$ in their results \citep{Asgari2019}. Even though such discrepancies could be attributed to a failure of the $\Lambda$CDM model \citep{Joudaki2016}, such a claim would need significant evidence and thorough testing. Some studies suggest it may instead be pointing to systematic biases in the weak lensing analysis methodologies \citep{Troxel2018a,Joudaki2019,Asgari2019,Wright2019}. Moreover, such studies indicate that a major difference in analysis methodologies lies in the redshift calibration, and that this can produce such discrepancy. Redshift calibration clearly needs substantial improvement for the current- and next-generation photometric surveys.      

Several techniques for estimating the redshift distributions of imaging surveys have been developed in the last decades, which can be broadly separated in three categories.
\begin{itemize}
\item
Direct \emph{spectroscopic} measurement of redshifts is an obvious tactic.  Since spectroscopic redshifts are expensive, this method must at present be applied to only a subset of the photometric galaxy population.  The resulting shot noise implies that $O(10^5)$ \emph{unbiased} spectra must be obtained \citep{MaBernstein2008}, to reach the needed precision in $n(z),$ potentially lowered to $O(10^4)$ by careful targeting to span the galaxy population \citep{Masters2015}.  Such ``direct'' calibration is also subject to large-scale-structure (LSS) variance if (as is most practical) it is conducted over a small sky area.  Even if numerical requirements are met, the direct method is subject to redshift biases because of differential rates of success in obtaining a reliable redshift across the redshift and magnitude range of the target population.  In this paper we will refer to information obtained through direct spectroscopy (or many-band photometry) as the ``spectroscopic prior.''  It is essential that any $n(z)$ estimation be robust to the noise and biases that exist in real-life spectroscopic surveys.
\item
\emph{Photometric} redshifts compare a set of observed fluxes (or
colors or potentially other measurable features) $F_i$ 
of source $i$ to those expected for galaxies at various 
redshifts to infer the redshift of the individual target galaxy. The map $z(F)$ is based on some mix of theoretical models of galaxy spectra with empirical knowledge from direct spectroscopy.  The inference $p(z|F)$ can be made
using explicit template-fitting methods~(e.g. {\tt Hyperz}, \citet{Bolzonella2000}; {\tt BPZ}, \citet{Benitez2000,Coe2006}; {\tt LePhare}, \citet{Arnouts2002,Ilbert2006}; {\tt EAZY}, \citet{Brammer2008}), or machine-learning ``training'' methods ~(e.g. {\tt ANNz}, \citet{Collister2004}; {\tt ArborZ}, \citet{Gerdes2010}; {\tt TPZ}, \citet{CarrascoKind2013a}; {\tt SkyNet}, \citet{Bonnett}).  The most basic, completely empirical, form of photometric redshift determination is to assign to each target the redshift of its nearest neighbor (by some metric in color/mag space) among a subset with spectroscopic redshifts.  This reweighting of the spectroscopy by imaging data was proposed by \citet{Lima2008}, and multiple current implementations of it attain various levels of rigor in the treatment of observational errors. 
Comparisons of different photometric methods have been performed in simulated and real data
\citep{Hildebrandt2010,Dahlen2013,Sanchez2014}.  The limitations of photometric methods are that the map $z(F)$ can be ambiguous even with noiseless data, therefore requiring that the correct $p(z|F)$ incorporate accurate priors on the relative abundance $n(z,F).$  And of course the photometric method inherits any biases or deficiencies of its theoretical/empirical training basis.
\item
\emph{Clustering} redshifts use data coming from large-area surveys 
to constrain $n(z)$, \ie~using the observed sky positions $\theta_i$ of the sources in comparison to the positions of a \emph{tracer} population with secure redshifts.  The simple principle is that the targets' $\theta_i$ will show no correlation with the tracers unless they are physically co-located, \ie~at a common redshift.  The tracers need not be a representative sampling of the targets. The weaknesses of this method is that it will, of course, only provide information at redshifts where abundant tracers are known; that the information per target is weak; and that the inference of $n(z)$ is degenerate with a redshift dependence of the ``bias'', \ie the relation between tracer space density and target space density.  As a consequence, the accuracy of the clustering method is enormously improved if photometric information can be used to select subpopulations known to have narrow redshift range.
The application of this method is typically based on 2-point statistics between the source population
and tracer population.
\citep{Newman2008,Menard2013,Schmidt2013}. 
\end{itemize}

Some recent analyses have attempted the combination of photometric and
clustering constraints on the same survey data in the presence of prior spectroscopic information
\citep{Hildebrandt2017,Hoyle2017,Gatti2018,Davis2017,DESCollaboration2017}, but the comparisons have been performed
just by means of basic visual cross-checks on the two independently
derived $n(z)$'s, or using
some single summary statistic of $n(z)$, such as its mean.

\citet{Sanchez2019}, hereafter SB19, present a framework to combine these three pieces of information (prior, photometry and clustering) in a principled way to assign a posterior probability to $n(z)$ using a hierarchical Bayesian model (see also \citealt{Leistedt2016}). The framework provides posterior samples of the redshift distribution of a population constrained by all sources of information, and SB19 demonstrated its performance on simple, idealized simulations. In this work, we extend the method so that it can be applied to real data from galaxy surveys, with the main addition being a practical, realistic treatment of the galaxy clustering information.
Using the public MICE2 N-body simulations, we define a galaxy subsample as tracers with known redshifts to develop a clustering probability based on a kernel density estimator (hereafter KDE). We incorporate a redshift-dependent \emph{biasing function} that maps the local tracer KDE output to the actual density distribution of the target galaxies. The method includes marginalization over the biasing functions' parameters, since in data there will be no sufficiently accurate prior on the biasing relations.

The methods developed in this work provide the necessary tools for the application of the framework to real data. The simulation that is used, even though it is not intended to completely mimic the real data, has all of the parameters relevant to the $n(z)$ accuracy in a realistic range, \ie\ galaxy and tracer density, clustering amplitudes and power spectra, noise levels, and sizes of spectroscopic, wide, and deep samples used in the application of the scheme to the \textit{Dark Energy Survey (DES)}. Therefore, the methods and the results presented in this paper demonstrate the capabilities of the framework in a realistic setting.   

This paper is organized as follows. In Section \ref{sec:framework} we present the details of the methodology and the phenotype approach. Section \ref{sec:sims} describes the simulated galaxy catalog used to test the methodology. We follow with a description of the density estimation used to incorporate the clustering information in Section \ref{sec:clustering}. We describe the Gibbs sampling technique used to sample the posterior on all the model parameters in Section \ref{sec:sampling}. Section \ref{sec:results} shows the main results in this work, with priors coming from precise redshifts over a small patch of sky.  We examine three cases in which these spectroscopic priors are biased, inspired by shortcomings in real data. Section \ref{sec:discussion} presents a discussion of the methodology and its application to real data, and we summarize and conclude in Section \ref{sec:conclusions}.

\section{Framework}
\label{sec:framework}


We work under the framework presented in SB19, in which galaxy ``types'' are defined by observed properties rather than rest-frame properties, and we call them \emph{phenotypes}.  The individual galaxies $i$ are seen as being drawn from a pool of possible phenotypes $t_i,$ redshifts $z_i,$ and angular positions $\theta_i,$ with intrinsic mean density $n(t,z)$ on the sky.  The $t_i$ and $z_i$ and $n(t,z)$ are noiseless latent variables, with the observations yielding a catalog with the $\theta_i$ and a noisy set of observable features which will be denoted as $F_i$---namely apparent magnitudes/colors. The clustering information is included by considering that the sky density of galaxies of type $t$ at redshift $z$ is modulated by some factor $1+\delta_z^t(\theta).$ In this paper we will simplify the galaxy density field to be type-independent, $\delta^t_z\rightarrow \delta_z.$  
The latent densities $\delta_z^t(\theta)$ will be constrained using a ``tracer'' galaxy population known to be at redshift $z.$
Our notation will be that the vector quantities \features, \types, \redshifts, and \positions\ denote the full set of properties of all selected galaxies, \ie\ $\features=\{F_1,F_2, \ldots, F_N\}$ (a summary of all the notation can be found in Table \ref{tab:notation}).

\subsection{Generative model}

As in SB19, the fundamental assumption of the method is that galaxies are drawn from a 
\textit{Cox process} \citep{Cox1955}
or \textit{doubly stochastic Poisson process}, \ie we assume that each galaxy is
Poisson sampled from a latent, stochastic density field. The problem simplifies when considering the redshift $z$ as an
integer indexing a set of finite-width redshift bins, where each bin has
an independent
density fluctuation field $\delta_z(\theta)$, \ie $\langle
\delta_{z_i}(\theta) \delta_{z_j}(\theta)\rangle = 0$ for $z_i\ne
z_j.$  We will also assume that we have a finite set of phenotypes indexed by integer $t$.
Each phenotype has a mean sky density of $n^t = n f_t$, where we place $n$ as the total density of all detectable galaxy phenotypes, and $f_t = p(t)$ being the fraction of the population in each type, with $\sum_t f_t=1$ as a constraint. Then the redshift distribution of type $t$ will be $p(z | t)=f^t_z$, and we will also denote
\begin{equation}
	\label{pzt}
f_{tz} \equiv p(z,t) =p(z | t) p(t) = f^t_z f_t.
\end{equation}

We are considering the sky to be populated with galaxies with a finite variety of redshifts and phenotypes, where phenotypes specifiy a galaxy's noiseless, observed appearance. We assume there is some selection function $s$ with the probability of a galaxy being selected, possibly depending on sky position, specified as a selection function $p(s|t,\theta).$ We will always assume that we know nothing about the non-selected galaxies, not even that they exist; the observed data are the positions \positions\ and features \features\ of the selected galaxies. All galaxies of phenotype $t$ observed under the same conditions
are assumed to have the same selection function $p(s | t, \theta)$ and the same probability
$p(F, s | t, \theta)$ of  being selected and measured to have image
features $F$. Finally, we will allow for some local biasing function, $\mathcal{B}^t_z$, with parameters $\biases_z^t$, depending on both redshift and phenotype, to relate the galaxies' spatial distribution to
the underlying tracer density fluctuation $\tilde\delta_z$. Now the selected galaxies can be considered as being a Poisson sampling of the following density field:
\begin{equation}
	\rho(z,\theta, t | n, \fractions, \biases, \tilde\fields) = n f_{tz} \mathcal{B}_z^t\left(\tilde\delta_z(\theta),\biases_z^t\right)p(s|t, \theta).
  \label{density2}
\end{equation}
The $\mathcal{B}$ term describes the spatial variation of the expected detection rate due to density fluctuations.  The last term describes density fluctuations due to variable observing conditions.  In this work we will consider the bias function to be independent of type, so $\mathcal{B}^t_z\rightarrow\mathcal{B}_z,$ and the biasing parameters likewise are independent of $t$.

With knowledge of the survey noise properties and the noiseless appearance of phenotype $t$, we can determine the likelihood $p(F, s | t, \theta, z)$ of a galaxy of phenotype $t$ at location $\theta,z$ being selected and measured with features $F$.  Note this likelihood will not depend on $z$ since the phenotype's observables are independent of $z$, by construction. Therefore, for each observed galaxy $i$ and phenotype $t$, we can assign a feature/selection likelihood 
\begin{equation}
  \likeli_{it} \equiv  p(F_i, s | t_i, \theta_i).
\end{equation}
This function will depend on the quality of the observations at sky position $\theta_i$ and the measurement and selection algorithms. We will assume that this likelihood is known \textit{a priori}, \eg\ by the result of analyzing the injection of artificial copies of the phenotype into the real survey images \citep{Suchyta2016}.

Then the probability of selecting a set of galaxies $i\in \{1 \ldots N\}$ at positions \positions\ with features \features, types \types\ and redshifts \redshifts\ takes the standard Poisson form:
\begin{align}
\label{full1}
p\left(\features, \positions, \types, \redshifts | n, \fractions,\biases, \tilde\fields\right)
  & = \exp\left[-n \sum_t f_t A^t(\fractions^t, \biases^t, \tilde\fields)\right] \\ 
  & \phantom{=} \times 
		    \prod_i \likeli_{it} n f_{t_i z_i}  \mathcal{B}^{t_i}_{z_i}\left(\tilde\delta_{z_i}(\theta_i),\biases_{z_i}\right) .
\nonumber 
\end{align}
The exponentiated quantity is, as required for Poisson distributions, the expected number of detections $\langle N \rangle$ for the entire sample. This can be determined from knowledge of the survey properties: 
\begin{align}
\nonumber
A^t(\fractions,\biases,\tilde\fields) & \equiv
      \sum_z  \int \mathrm{d}^2\theta\, p(s | t, \theta) f^t_z\mathcal{B}^t_z\left(\tilde\delta_{z}(\theta),\biases_{z}\right) \\
\nonumber
 & = \int \mathrm{d}^2\theta\, p(s | t, \theta) \sum_z   f^t_z \mathcal{B}^t_z\left(\tilde\delta_{z}(\theta),\biases_{z}\right)\\
 & \approx \int \mathrm{d}^2\theta\, p(s | t, \theta),
\label{aeff}
\end{align}
where we have assumed that the clustering information integrated over the mask of the survey approximately keeps its average value of unity, $\int \mathrm{d}^2\theta\, p(s | t, \theta) \mathcal{B}^t_z\left(\tilde\delta_{z}(\theta),\biases_{z}\right) \approx 1$.   

In order to provide the full generative model for the data we must also specify
the process $p(\tilde\fields | \pi_\delta)$ generating the stochastic density
fluctuation fields given some hyperparameters $\pi_\delta$. For instance, that could be
a log-normal distribution where $\pi_\delta$ specifies the power spectrum.  We also require priors $p(\biases)$ and $p(n)$, plus any prior information on $p(\fractions)$ aside from the constraint that $\sum_{tz} f_{tz}=1.$  

\begin{table}
	  \caption{Summary of the notation used throughout this paper.}
    \label{tab:notation}
    \begin{center}
    \begin{tabular}{cl} 
      \hline
	    $F$ & galaxy set of observed features \\
	    $t$ & galaxy phenotype (or simply type) \\
	    $z$ & galaxy redshift \\
            $\theta$ & galaxy angular position in the sky \\
            $s$ & indicator of successful detection/selection \\
	    $\likeli_{it}$ & probability of measuring galaxy $i$ with $F_i$ given $t$ \\
	    $\features, \types, \redshifts, \positions$ & set of properties for all galaxies in the sample  \\
	    $N$ & number of galaxies in the sample \\
	    $N_t$ & number of types \\
	    $N_z$ & number of redshifts \\
	    $A$ & effective area of the survey for source detection \\
	    $n$ & mean galaxy density per unit solid angle \\
	    $n(z)$ & mean galaxy density per unit solid angle per $z$ \\
	    $\delta_z(\theta)$ & target density fluctuation at a given $z$ and $\theta$ \\
	    $\tilde\delta_z(\theta)$ & tracer density fluctuation at a given $z$ and $\theta$ \\
	    $\hat\delta_z(\theta)$ & tracer density estimator at a given $z$ and $\theta$ \\
	    $\pi_\delta$ & density fluctuation field hyperparameters \\
	    $\fields$ & set of $\delta_z(\theta)$ for all redshifts and  positions \\
	    $\mathcal{B}$ & mapping relation between estimated\\
	    & density field and true clustering probability \\
	    $b_z^t$ & parameters of the $\mathcal{B}$ function for type $t$ at redshift $z$ \\
	    $\biases$ & set of $b_z^t$ for all types and redshifts \\
	    $f_{zt}$ & joint type and redshift probability $p(z,t)$  \\
	    $\fractions$ & set of $f_{zt}$ for all types and redshifts \\
	    $\nzt$ & number of sources assigned to redshift $z$ and type $t$ \\
	    $\counts$ & set of $\nzt$ for all redshifts and types \\
	    $\mzt$ & number of sources in the prior at redshift $z$ and type $t$ \\
	    $\priorcounts$ & set of $\mzt$ for all redshifts and types \\
	    $\Delta z$ & difference between the means of \\ 
	     & estimated and true $n(z)$'s \\
	    $D_{\mathrm{KL}}$ & Kullback-Leiber divergence between\\
	    &  estimated and true $n(z)$'s\\
      \hline
    \end{tabular}
    \end{center}
\end{table}

\subsection{Redshift inference}
\label{sec:sompz}
The principal quantity of interest is the underlying redshift distribution
\begin{equation}
n(z) = n \sum_t f_{tz}.
\end{equation} 
In most applications of redshift inference we are only concerned with the shape, not the normalization, of $n(z)$, and therefore we will focus here on the fractions \fractions, rather than $n$. In addition, in many applications it is also useful to know the individual redshifts of galaxies \redshifts, and in order to enable a Gibbs sampling scheme, which is the simplest way of sampling our posterior (see Section \ref{sec:sampling}), we will need to keep \biases\ and \types\ as conditional variables. We can use Bayes' theorem to write down the posterior joint probability of these variables of interest: 
\begin{align}
\label{pfz1}
p(\fractions, \redshifts, \biases, \types | \features, \positions, \pi_\delta) & \propto
 \int \mathrm{d}n \,  \mathrm{d}\tilde\fields  \\
  & \phantom{=\int} 
  p\left(\features, \positions, \types, \redshifts | n, \fractions, \biases, \tilde\fields\right)  p(\tilde\fields | \pi_\delta)  \, p(n) \, p(\fractions) \, p(\biases).
\nonumber
\end{align}
We have already derived the first term under the integral in \eqq{full1}. In this paper, as in SB19, we will work with the approximation that we can replace the stochastic tracer density fluctuation $\tilde\delta_z(\theta)$ with some deterministic estimator $\hat\delta_z(\theta)$ of the realization of the density fields in the generative probability of \eqq{full1}. Under that approximation we can ignore the hyperparameters generating the density field $\pi_\delta$ but we lose the ability to use the information available from the clustering of the target galaxies. Performing the marginalization over $n$ assuming the effective area of the survey is independent of phenotype (see SB19 for more details), the posterior distribution for redshift and phenotype information in \eqq{pfz1} becomes
\begin{align}
\label{pzf3}
p(\fractions, \redshifts, \types, \biases | \features, \positions) & \propto
p(\fractions) p(\biases) \,  
	\prod_i  \likeli_{it_i} \, f_{t_iz_i} \, \mathcal{B}_{z_i}\left(\hat\delta_{iz_i},\biases_{z_i}\right), \\
\hat\delta_{iz} & \equiv \hat{\delta}_z(\theta_i). \end{align}

The roles of the main three sources of information in redshift estimation are clearly present and differentiated in the posterior of \eqq{pzf3}.  First, there is a term for the prior probability that any galaxy is of phenotype $t$ and redshift $z$, $f_{tz}$. Second, the photometric information for a galaxy is in $\likeli_{it}$, which is the likelihood of galaxy $i$ resembling phenotype $t$ and passing selection. Third, clustering information enters as the last term, describing the modification of the probability by (our estimator for) the density fluctuation field.  

In more detail: the prior term can be estimated using a subset of galaxies with well characterized phenotypes and redshifts, which will call the \textit{spectroscopic} sample. It requires deep (low-noise) photometric data, plus either spectroscopic or high-quality photometric redshifts, of a fair subsample of the sources. The clustering information will require another galaxy sub-population, the \textit{tracers}, having well-characterized redshift information and spanning a large area and redshift  range of the survey (but no need to span them completely). This can be a population of galaxies with accurate photometric redshift estimates, like \eg~luminous red galaxies (LRGs). We will refer to all galaxies in the sample of interest as \textit{target} galaxies, for which we will only have the measurements of \features and \positions.             


\subsection{Realistic set up: SOM implementation}

To discretize the phenotypes for a general imaging survey, we propose to use a combination of wide and deep survey observations and \textit{self-organizing maps} (SOMs, \citealt{Masters2015}). Deep observations are often available for surveys like the \des\ by summing observations of fields being monitored for high-$z$ SNe.  These provide essentially noiseless photometric measurements and observations in additional filter bands for galaxies in specific fields (henceforth deep fields, or simply DFs).  The DFs provide an empirical sampling of the distribution of galaxies in feature ($\features$) space. In turn, SOMs provide a data-driven way of mapping and discretizing that feature space, so that each cell $c$ of the so-called \textit{deep} SOM cell constitutes a phenotype $t$.  

Another term that we will need in the data application is the noise or measurement likelihood, $\likeli_{it} \equiv  p(F_i, s | t_i, \theta_i)$. We follow the approach of \citet{Buchs2019} and construct the measurement likelihood by training another SOM on wide-field data of the galaxy survey of interest; we will refer to this one as the \textit{wide} SOM and its cells, $\hat{c}$, span the space of features \features\ observed in the wide-field survey (\ie\ every detected galaxy will be assigned to one \textit{wide} cell, $\hat{c}$). Crucially, it is possible to inject artificial copies of galaxies with deep photometry, and hence well specified phenotypes, into the real images of the survey, and measure their (noisy) wide-field properties \citep{Suchyta2016}. Then, for a set of injected galaxies, we will know both the cells in the deep and wide SOMs ($\hat{c}$ and $c$), so that we can construct the mapping between deep and wide SOMs which corresponds to our measurement likelihood:
\begin{equation}
	\likeli_{it} \equiv  p(F_i, s | t_i, \theta_i) \equiv p(\hat{c}_i, s | c_i, \theta_i).
\end{equation}

One other major part in the application of the method to data is the addition of clustering information, that is, the construction of the density field estimator using a tracer population and the creation of biasing functions $\mathcal{B}^t_z$ relating that estimate to the true underlying density fluctuation field of the selected galaxies. This will be treated in Section~\ref{sec:clustering}.

\section{Simulations}
\label{sec:sims}

SB19 demonstrated the HBM method on a simplified simulation with idealized galaxy properties and noise distributions, and perfect knowledge of the density fluctuation field. Now, instead, we test our methodology on the public MICE2 simulation,\footnote{The data can be downloaded from CosmoHub \citep{COSMOHUB}, \url{https://cosmohub.pic.es/}.} a mock galaxy catalog created from a lightcone of a dark-matter-only N-body simulation that contains  $\sim$200 million galaxies over one sky octant ($\sim 5000$ deg$^{2}$) and up to $z=1.4$. Several important differences with respect to the SB19 simulation make this analysis more realistic and allow the method described herein to be applicable to analysis of real data.   

First, the MICE2 simulation has realistic clustering properties given by a $\Lambda$CDM cosmology with parameters $\Omega_{m}=0.25$, $\Omega_b=0.044$, $h=0.7$, $n_s=0.95$, $\Omega_{\Lambda}=0.75$, $\sigma_8=0.8$ and $w=-1$. In addition, we do not assume true knowledge of the density field but rather infer the clustering information from a set of galaxy tracers, described below. Second, galaxies have realistic spectral energy distributions (SEDs) assigned from the COSMOS catalog (\citealt{COSMOSILBERT}) that reproduce the observed color-magnitude distribution as well as clustering observations as a function of colors and luminosity (see \citealt{MICE2} for more details). Once the galaxy SED is known, mangnitudes are computed based on the luminosity and redshift of the galaxy. The galaxy properties, clustering and lensing in the simulation have been thoroughly validated in \cite{MICE0,MICE1,MICE2,MICE3}.

\subsection{Target and tracer sample selection}
\label{sec:sample}

We select a galaxy sample within a square footprint defined by the cuts $30\leq \mathrm{RA[deg]}\leq 60$ and $0\leq \mathrm{Dec[deg]}\leq 35$, representing an area of around 1000 sq.~deg., with the redshift range $0.2\leq z \leq 1.2$ and a magnitude limit  $i_{\mathrm{DES}}<24$. To reduce runtimes, we cull the galaxy catalog by a factor $\sim 2$ by selecting only those galaxies with a subset of SEDs.
This downsampling retains a representative sampling of populations (Elliptical, Spiral, Starburst) and dust attenuation laws and values present in the simulation.\footnote{The selection is defined as sed\_cos $\equiv c \in 0,1,2,5,6,7,10,11,12,$\\$15,16,17,21,22,23,24,25,29,30,35,36,37,38,39,41,42,43$.} 

\begin{figure}
\centering
\includegraphics[width=0.49\textwidth]{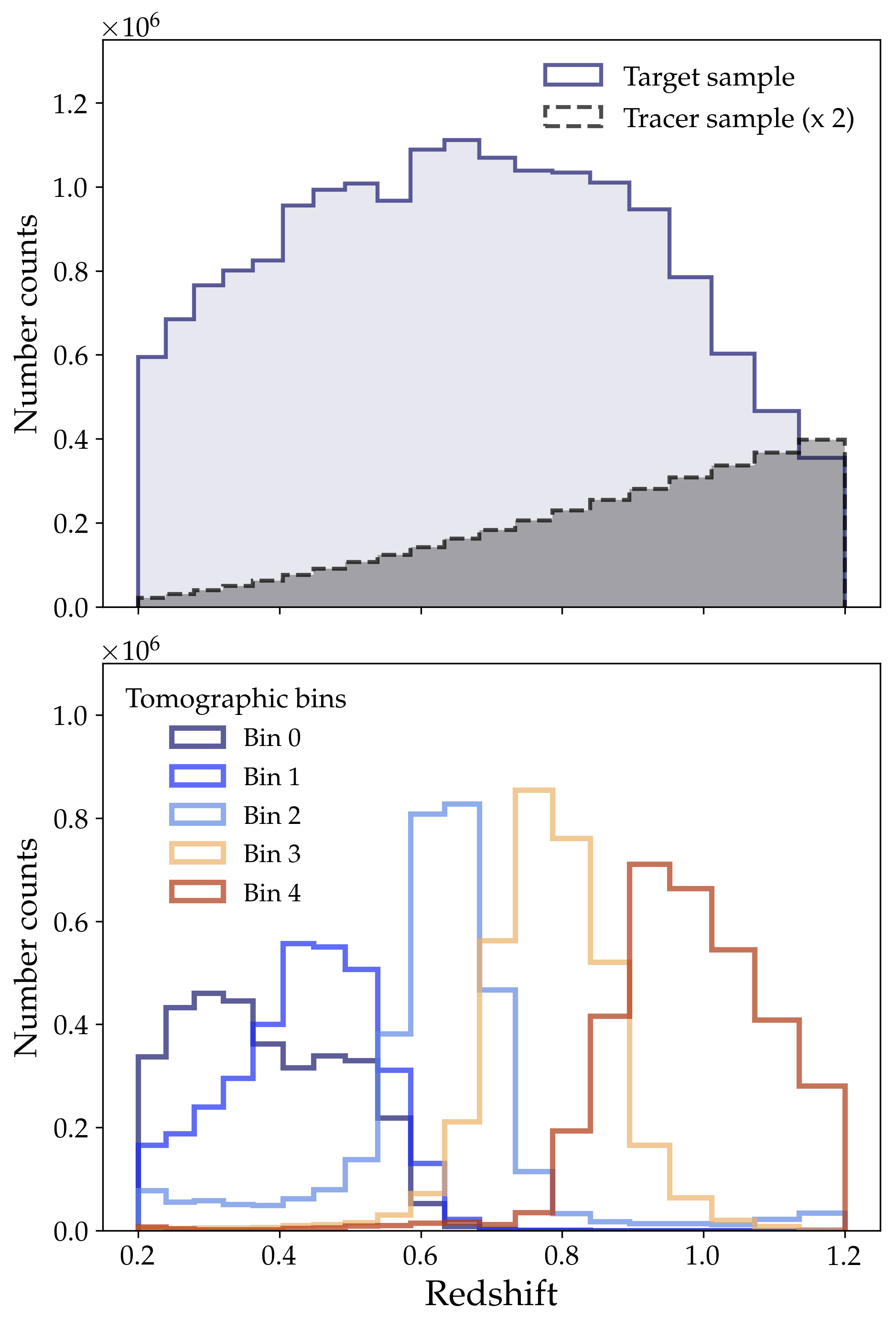}
	\caption{\textit{(Upper panel:)} Redshift distributions of the target and tracer samples. The target sample contains the galaxies for which we want to find a redshift distribution. The tracer sample contains galaxies with known redshifts that are used to add the clustering information into the redshift estimation. \textit{(Lower panel:)} Redshift distribution of tomographic bins defined as in \S\ref{sec:tomo}.}
\label{target_tracer_zdist}
\end{figure}

The tracer sample is a subsample of the full population, randomly drawn to maintain a constant comoving density similar to that of the RedMagic \des\ Y1 galaxy sample in its first three lens bins (\citealt{desy1_jackpaper}). This choice is arbitrary, and perhaps unrealistic at the higher end of our redshift range, but it is not a necessary feature of the method.
The target sample is defined as the galaxies that are not selected as tracers. The upper panel of Fig.~\ref{target_tracer_zdist} shows the redshift distributions of both samples. Tracers have a density between 0.015 (at $z=0.2$) and 0.5 (at $z=1.2$) times the target density. The redshift binning is chosen to have 20 bins equally spaced in comoving distance $\chi$ between the redshift limits of the catalog, which makes the tracer sample have a constant density per bin per unit comoving surface area $dA$, $dN\propto dAd\chi \propto dA$.  

It is worth highlighting here the differences between this simulation and a corresponding real data sample, in particular \des. First, the simulation sample used in this work contains about 1/5 of the total area in \des. This is relevant as we expect the clustering information to grow more powerful as area grows, so the simulation is a conservative estimate of the value of clustering. Second, the galaxy tracers used in adding clustering information in this work are unbiased with respect to the total sample. A real data application is likely to use luminous red galaxies (LRGs) or other highly biased population as tracers.
We have not, however, \emph{assumed} in this simulation that tracers are unbiased, but we have instead marginalized over a biasing relation.
The lower tracer bias (relative to mass) in our simulation is a conservative scenario, since it will increase the impact of shot noise in the density estimates compared to an LRG tracer sample. Finally, we have used a limited redshift range in this work, $0.2 < z < 1.2$, and we have used tracers spanning this entire redshift range. In the application to real data, a more complete redshift range will have to be considered, and tracers may be available just for a limited redshift range, but that can be accommodated naturally in the method and was shown to work as expected in SB19.         


\begin{figure*}
\centering
	\includegraphics[width=\textwidth]{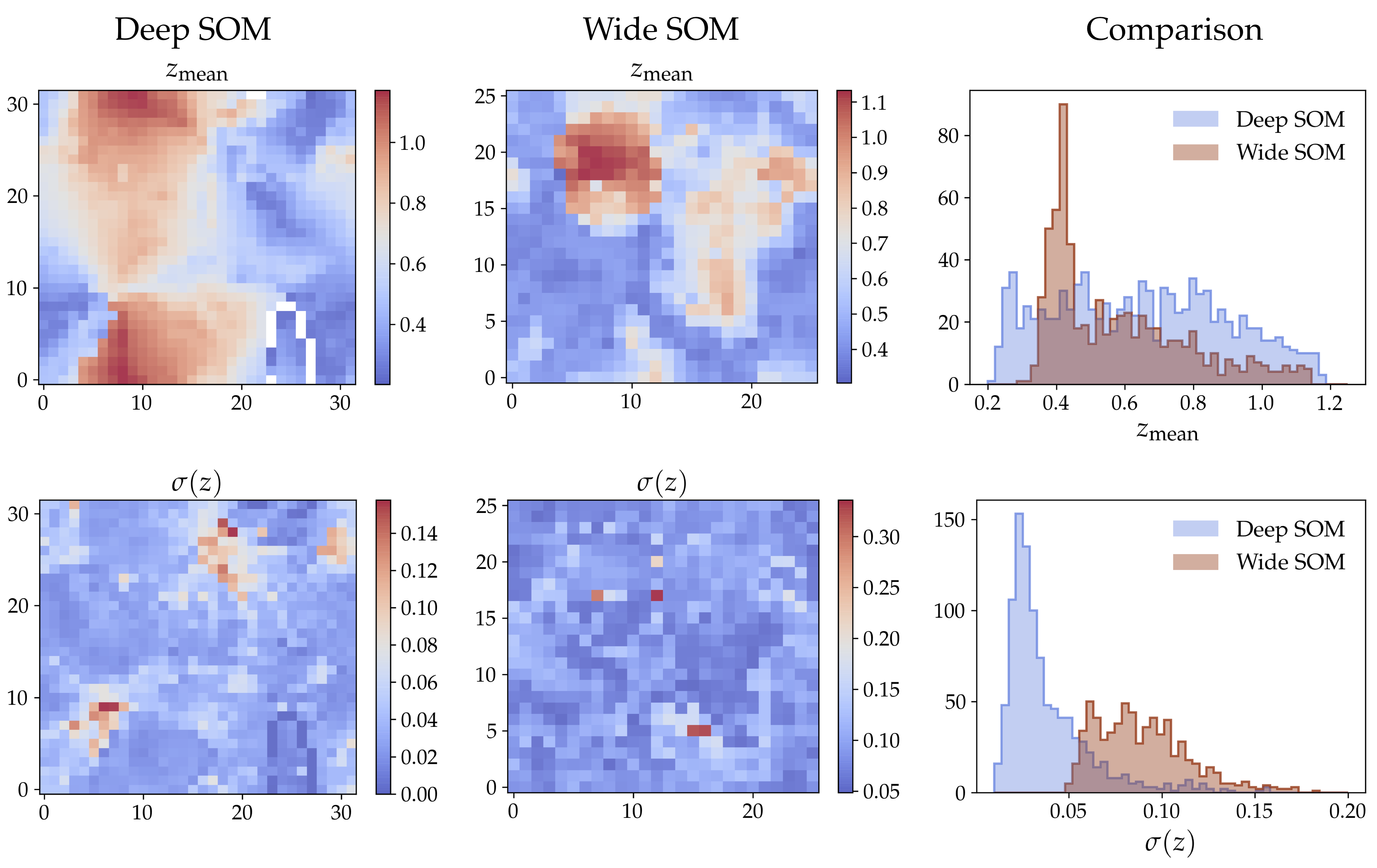}
	\caption{Mean redshift and redshift dispersion of cells in deep and wide SOMs described in \S\ref{sec:soms}. The left and central columns show the SOM maps populated with these quantities, while the plots in the right column show the comparison of these distributions. These show how the deep SOM better samples the redshift space of the simulation test, with a lower redshift scatter per cell. }
\label{deep_som}
\end{figure*}

\subsection{The phenotype approach: Deep and wide SOMs}
\label{sec:soms}

The phenotype method described in Section \ref{sec:framework} is then applied to the simulation. As stated in \S\ref{sec:sompz}, the approach can benefit from a \emph{deep} sample with deeper photometry and extra observed wavelength bands than the target (or ``wide'') sample, which helps define galaxy phenotypes that individually span narrower redshift ranges. We choose among the available bands in MICE2 the \des\ $g$,$r$,$i$,$z$ bands for both samples, and the additional CFHT $u$, \des\ $Y$ and VHS $J$,$H$,$K$ bands for the deep sample, mimicking the \des\ wide and deep survey fields. For the deep sample, we stick to noiseless true fluxes from the simulation, while for the wide sample we add Gaussian noise to the fluxes by fitting a linear relation between magnitude and logarithmic magnitude error for each band using observed noise from the \des\ Year 1 public data\footnote{\url{https://des.ncsa.illinois.edu/releases/y1a1/key-catalogs/key-mof}}. We produce deep and wide photometries for all galaxies of the target sample. We finally select only galaxies that have a signal to noise above 5 in each wide band, $g,r,i,z$.\footnote{Before adding the noise, we shift each galaxy's magnitude by $-1.2$, to increase the number density of our target sample passing $S/N$ cuts to 4.7 galaxies/$\mathrm{arcmin}^2$, as observed in the \des\ Y1 Metacalibration source sample  (\citealt{Troxel2017}). This counteracts the MICE catalog culling described in Section~\ref{sec:sample}.}
We leave for future work an accurate abundance matching of the color and magnitude distributions of the simulation to the observed ones from real galaxy survey.

Following \S\ref{sec:sompz}, we create two self-organizing maps (SOMs) on square grids with periodic boundary conditions, each similar to the SOM in \citet{Masters2015}. The deep SOM is trained with eight colors, defined as $\mathrm{mag}-i$, where $\mathrm{mag}=\{u,g,r,z,Y,J,H,K\}$, in a $32\times32$ grid. The wide SOM is trained with one magnitude, $i$, and three colors, $g-i$, $r-i$ and $z-i$, in a $26\times26$ grid. Each color is renormalized to span the range $[0,1]$, while the magnitude spans the range $[0,0.1]$, \ie, we give colors $10\times$ more weight than magnitude in creating the wide SOM. Also, to avoid noise influencing the training of the wide SOM, we only use galaxies with a $S/N>10$ to do so.
In the simulation, we know the truth and an observed magnitude for every galaxy in the target sample, so we can assign a $\hat c$ from the wide SOM and a $c$ from the deep SOM to every galaxy (the ``best-matching unit'', or BMU, in SOM parlance).  From these we can
calibrate the probability $p(\hat{c}|c)$. In the application of this method to real data, the ``truth'' (low-noise fluxes) are not available for every target but only to a subset, so that only a wide-SOM assignment $\hat c$ is available for all targets.  But $p(\hat c | c)$ can be estimated through repeated injection of deep-sample galaxies into the wide images, serving the same purpose.  Both methods should yield an accurate assessment of $p(\hat c, c),$ which is essential for success of any photometric approach to redshift estimation from noisy fluxes.

Figure \ref{deep_som} shows the mean redshift and redshift dispersion of the cells in the deep and wide SOMs described above (left and central columns). From the plots, one can note the smoother redshift distribution and the lower redshift dispersion in the deep SOM compared to the wide SOM. This is even more evident from the comparison plots in the right column of Fig.~\ref{deep_som}: the distribution of the mean redshift per cell in the deep SOM is more uniform and samples better the redshift space of the simulation ($0.2 < z < 1.2$), and the redshift dispersion per cell in the deep SOM is significantly lower (median $\sigma(z)$ of 0.030 for the deep SOM vs. 0.086 for the wide SOM).  

\subsection{Tomographic bins}
\label{sec:tomo}

Tomographic redshift bins are defined as groups of wide-SOM cells.  We first find the mean expected redshift for each wide cell as 
\begin{equation}
z_{\mathrm{mean}}(\hat c)=\int dz \, z\ \left[\sum_c p(z|c) p(c|\hat{c})\right], 
\end{equation}
where $p(z|c)$ is also estimated using all galaxies in the target sample. We sort the wide SOM cells by their $z_{\mathrm{mean}}$, then split them into 5 contiguous redshift bins with equal number of galaxies. The true redshift distribution of each tomographic bin is shown in the lower panel of Figure~\ref{target_tracer_zdist}. 
The estimation of $n(z)$ presented in this work can be applied to any subset of the target galaxies defined by the features $F_i$, but here we will use an example the determination of $n(z)$ for Bin 3 of this scheme.
To do so, we first we retrain the wide and deep SOMs using only those target galaxies whose noisy photometry places them into this bin.  This choice potentially avoids some biases that can arise from differential bin selection within the finite range of redshifts in individual deep-SOM cells, as highlighted by \citet{Wright2019} and other works.  This step can be executed on real data using the deep sample.

\subsection{Spectroscopic sample}
To determine a prior $p(t,z)$ we will make use of a spectroscopic sample for which both $t$ and $z$ are assumed to be known definitively for each galaxy passing target-sample cuts.  In the simulation, the truth values are known exactly; 
in reality, they will typically come from a spectroscopic or high-quality photo-$z$ sample, and span a small area of the sky and are hence subject to sample variance. They are intended to be representative of the full galaxy population, but can be subject to incompleteness and biases. Our simulated spectroscopic sample consists of all target galaxies from one healpy sky pixel of the simulation (with nside=$2^5$), which has an area of $\sim3.5\mathrm{deg}^2$.
The same tomographic bin selection as made on the target sample is applied to the noisy versions of the photometry for the spectroscopic galaxies, leaving around 11,000 objects having spectra, in comparison to $3.3\times10^6$ galaxies in this bin from the full 1000~deg$^2$ target sample.

In Section~\ref{sec:results} we will investigate sample variance by choosing different regions for the spectroscopic sample, and also investigate the effects of placing measurement biases on the redshifts assumed for this sample.

\section{Adding the clustering information}
\label{sec:clustering}

As described in Section \ref{sec:framework}, we will work under the approximation that we can replace the latent density field of the tracer population with a set of deterministic estimators $\hat \delta_z(\theta)$ discretized in redshift space.  We also assume that these tracers are drawn from the same generative model as the targets, up to some local biasing relation $\mathcal{B}$ with parameters $b$, so that we are assuming $p(\theta | z)\propto \mathcal{B}_z[\hat\delta_z(\theta), b_z]$.

Before proceeding to describe the density estimators and biasing functions used in this simulation, we pause to note that we do not require the resultant $p(\theta | z)$ to be perfect or unbiased.  The correlation redshift method uses the density estimator $p(\theta | z)$ to inform us whether galaxies are more likely to truly be at $z$ than to be at some $z^\prime \ne z$.  In the latter case the target galaxies are distributed essentially randomly in $\theta$ with respect to $p(\theta | z).$  A useful figure of merit (FoM) for our density estimator is therefore the mean boost in (log) likelihood that a galaxy gets if it is assigned to its true redshift:
\begin{align}
  \mathrm{FoM}_{z} & = \left\langle{ \log p(\theta_i | z) }\right \rangle_{i \in z} - \left\langle{ \log p(\theta_i|z) }\right \rangle_{i \notin z} \\
  & = \left\langle  \log \mathcal{B}_z\left[\hat\delta_z(\theta_i),b_z\right] \right\rangle_{i \in z} - 
         \left\langle  \log \mathcal{B}_z\left[\hat\delta_z(\theta_i),b_z\right] \right\rangle_{i \in r},
\label{eq:fom}
\end{align}
where the first term is evaluated for galaxies truly at $z$, and the second term is for a population of galaxies randomly distributed across the footprint.  In the simulations we can evaluate this FoM over the full footprint, as a guide to good choices to make for the KDE and bias parameters.  In real data, this estimation is possible only over the smaller spectroscopic sample.

\subsection{Density estimation}
The tracer population, described in \ref{sec:sample}, is split in 20 redshift bins equally spaced in comoving distance in the range $z\in[0.2,1.2]$ using the true redshift from the simulation. The redshift bins are wider than the typical RMS redshift uncertainty of photometric LRGs in \des, which have, $\sigma_z\sim0.015(1+z)$ \citep{Rozo2015,KIDS_LRG}, and also wide enough to make their projected density fields nearly independent from each other. We will defer to future work any attempt to include photo-z errors in the tracer sample.

Several methods exist to reconstruct the surface density of galaxies (see \eg~\citealt{DTFE,Darvish}) from a point sample. In this work we will use a KDE to estimate the density field at any position of the field, using a circular kernel function $K(r)$:
\begin{equation}
\hat{\delta}_{z}(x) \equiv \frac{\frac{1}{N_{T}}\sum_{T}~K(\theta_{xT})}{\frac{1}{N_{R}}\sum_{R}~K(\theta_{xR})} -1.
\end{equation}
Here $\theta_{xT}$ runs over the distances between our sample point $x$ and each of the $N_T$ tracers at redshift $z$, while $\theta_{xR}$ runs over the pairs with a random sample of size $N_R$ that describes the selection function of the tracer sample.  The KDE is seen to be equivalent to 
the weighted two-point functions used in conventional $\mathrm{clustering-}z$ redshift techniques. 
We presume $N_R\gg N_T$ such that the dividing term can be considered a measure of the area surrounding $x$, taking into account the selection function and mask effects.

\begin{figure*}
\centering
\includegraphics[width=0.9\textwidth]{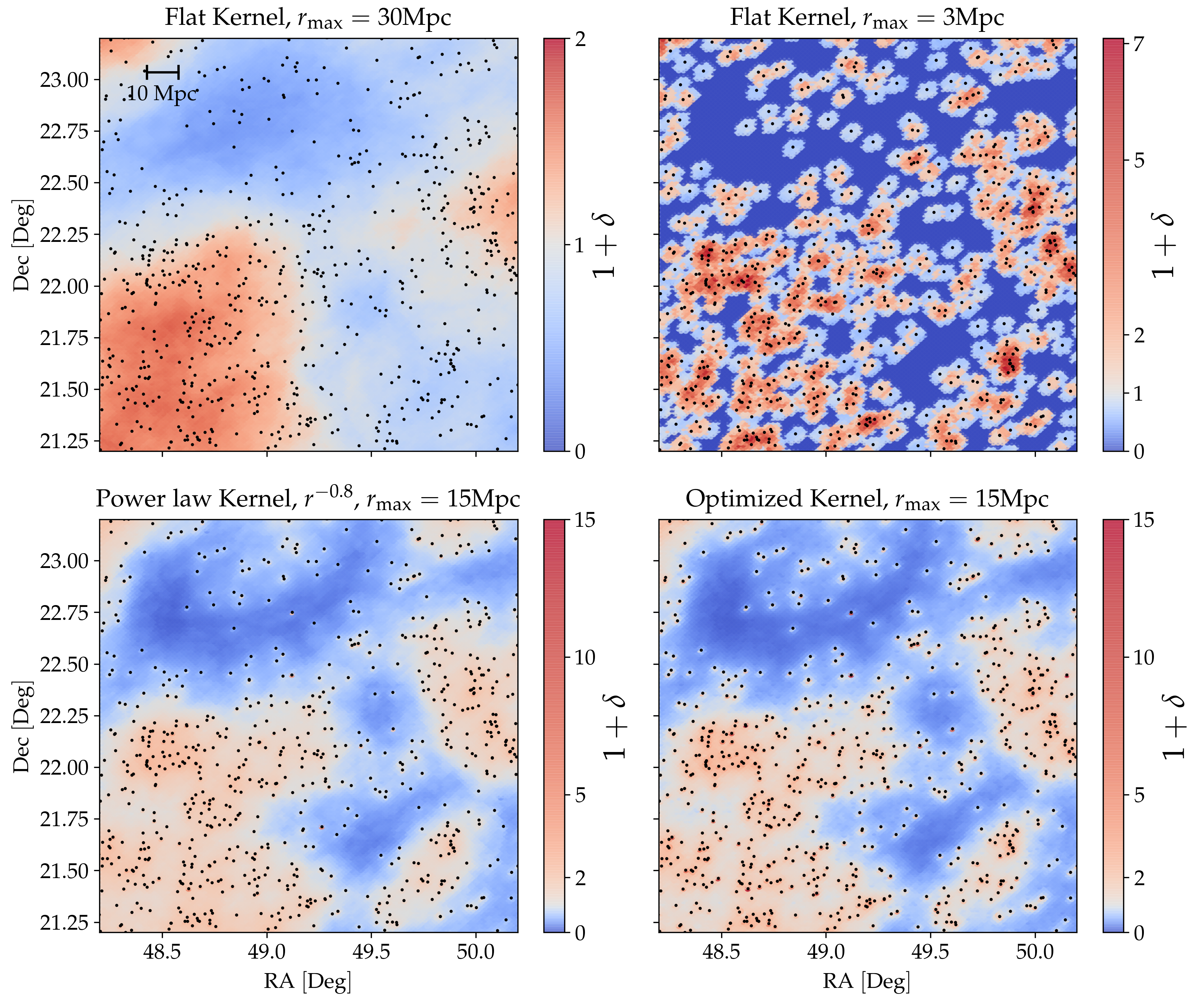}
\caption{Density field estimation using different kernel density estimators from a tracer sample population. Shows the field estimate for a small patch in the highest redshift bin. The black dots show the position of the tracer galaxies, and the background colors show the estimated value of the density field at different positions. The top panels show a flat kernel with a large size ($r_{\mathrm{max}}=30\mathrm{Mpc}$, left) and a small size ($r_{\mathrm{max}}=3\mathrm{Mpc}$, right). The bottom left panel shows the density with a power-law kernel that better resolves the structures. The bottom right panel shows a field estimated with an optimized kernel, which is our default density field estimate. Note the change in color scales in different panels, with white always corresponding to the mean density.}
\label{density_field_kde}
\end{figure*}

Choosing the shape and extent of the kernel $K$ is important. Figure \ref{density_field_kde} shows the effect that different kernel shapes have on the field estimate. The top left panel shows a top-hat kernel of size $r_{\mathrm{max}}=30\mathrm{Mpc}$. Such a large kernel smooths the density field too much and cannot resolve massive structures well, underestimating the density in cluster regions. The top right panel shows a small top-hat KDE with $r_{\mathrm{max}}=3\mathrm{Mpc}$. This KDE can better resolve dense structures, although it will still underestimate high-density regions, is more affected by shot noise, and indicates zero density in a large fraction of the sky.  SB19 show that, in simplified limits, the most informative kernel will match the angular correlation function of the galaxies, so that $K\propto r^{-0.8}.$

Many cosmological applications of redshift inference will also use statistics of the tracer sample as part of their constraining data. Allowing large scales into the KDE can improve its estimation, but will also correlate our resultant $n(z)$ with the observables being used for cosmology, which will complicate the derivation of cosmological parameter constraints.  Yet using only very small scales ($<3\,\mathrm{Mpc}$)  lowers the $S/N$ of the density estimator and the accuracy of $n(z)$ inferences.  We compromise by using a kernel that is zero for $r > r_{\mathrm{max}}=15\,\mathrm{Mpc}$, although we also explore $r_{\mathrm{max}}=10\,\mathrm{Mpc}$ for comparison.  Note that \des\ cosmological analyses use correlations only above 8--12~Mpc.
The bottom left panel shows a power law  $K(r)\propto r^{-0.8}$, truncated at $15\,\mathrm{Mpc}.$

\subsection{Biasing relation}
In the simulation, we can calculate the true relation between galaxy density at some redshift $z$ and the KDE estimator $\hat\delta_z(\theta)$ by calculating the true source density:
\begin{equation} \label{eq:mapping_noparam}
\mathcal{B}_z^{\rm true}
(\hat\delta) \equiv \frac{\frac{1}{N_{T}} ~n_T(\hat\delta)}{\frac{1}{N_{R}}~n_R(\hat\delta)},
\end{equation}
where $n_T(\hat\delta)$ and $n_R(\hat\delta)$ are the number of galaxies and randoms in sky regions with some (small range of) KDE value.

Figure \ref{mapping_powerlaw_10mpc} shows, for each redshift bin (color coded), the relation between this average density of targets as a function of KDE value with a power-law KDE with $r_{\mathrm{max}}=10\mathrm{Mpc}$.
If the KDE delivered a perfectly unbiased field estimation, this would yield the dashed line. In general, the KDE estimate will not deliver such an estimate, both because the KDE yields a biased estimate of tracer density, and because the tracer will be a biased tracer of the target galaxies.  There is always a $\mathcal{B}^{\rm true}$ which will optimize the performance of a given KDE.  In real data we will not know this function in advance, so we 
propose a parametric form for the true probability $p(\theta|z)$ of a target galaxy being at position $\theta_i$ and redshift $z$,
\begin{equation} \label{eq:mapping_param_form}
p(\theta_i|z) = \mathcal{B}\left(\hat{\delta}_z(\theta_i), \{b^{z}_k\}\right),
\end{equation}
where $\{b^{z}_k\}$ are the parameters of $\mathcal{B}$ at redshift $z$. This is an approximation of a more general approach where the density field is updated locally by the targets as part of the hierarchical model. The parameters $\{b^{z}_i\}$ are part of the framework parameters (see Section \ref{sec:framework}) and they will be sampled along with the other parameters in the HBM (see Section \ref{sec:sampling}). We choose a polynomial of degree four as our mapping function $\mathcal{B}$, such that
 \begin{equation}
	 \label{eq:biasing}
\log_{10}(p(\theta|z)) = \log_{10}\mathcal{B}_z\left[\hat\delta_z(\theta)\right]=\sum_{k=0}^{4}~ b_k^{z}  ~\log_{10}(1.1+\hat{\delta}_z)^{k},
\end{equation}
with the additional constraints that $\int p(\theta|z) d\theta =1$ and that the derivative must always be positive. Note the use of $(1.1+\hat{\delta}_z)$ on the right-hand side to avoid singularities when the KDE yields $\hat\delta=-1.$

The use of a parametric biasing function adds another criterion to the choice of KDE kernel, because we will prefer a kernel which yields a more linear, less complex biasing function which we can expect to require fewer parameters and less variation with redshift.  These characteristics will improve our ability to fit optimal biasing functions to the KDE output.

While the biasing relation in general depends on both redshift and phenotype (see Section \ref{sec:framework}), we are neglecting the phenotype dependence throughout this work. The redshift determination could potentially be improved by, for example, allowing red galaxies a distinct bias from blue galaxies.  There will be potential degradation, though, as more free parameters are introduced into the model.  We defer an attempt at using this information for a future work.

\begin{figure}
\centering
\includegraphics[width=0.49\textwidth]{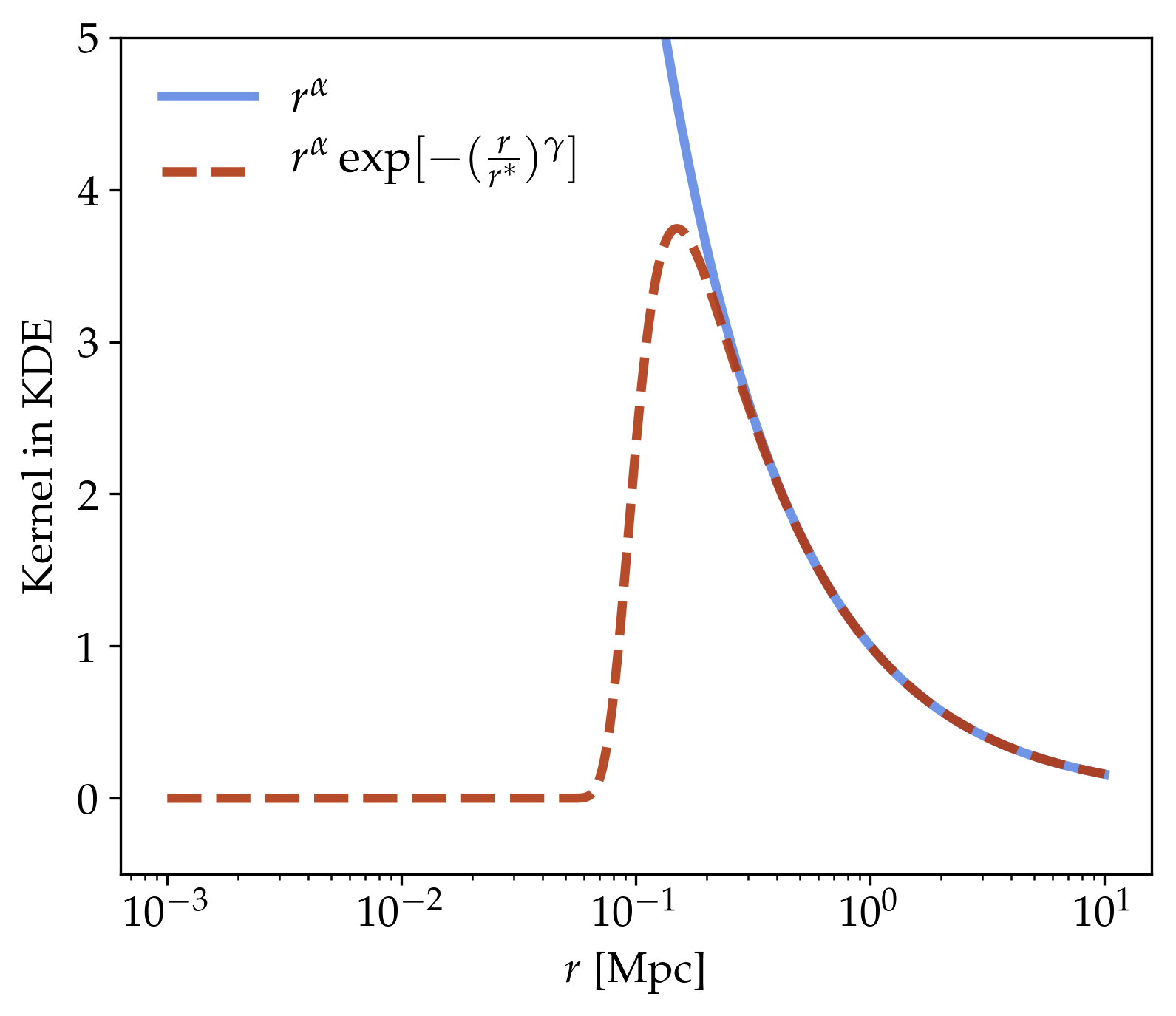}
\caption{Comparison between a power law KDE and a KDE with a power law that truncates at some scale $r^{*}$. Such truncation reduces the impact of shot noise in smaller scales and naturally adds a small exclusion region around the positions of tracers.}
\label{kde_optimize_formula}
\end{figure}

\subsection{Optimizing the estimator}

We can go one step further and try to optimize the shape of the KDE kernel, assuming we have a small calibration patch where the redshifts of the target galaxies are known. For this purpose we define a KDE with shape
\begin{equation}
\mathrm{KDE} \propto r^{\alpha}\exp\left[-\left(\frac{r}{r^{*}}\right)^{\gamma}\right], 
\end{equation}
%
which combines a power law with exponent $\alpha$ and an exponential truncation of the power law at scale $r^{*}$ with width parameter $\gamma$. Figure \ref{kde_optimize_formula} compares this kernel shape to a power law. The motivation for allowing a truncation at small scales is to reduce the effect of shot noise for sparse
tracer samples. 

\begin{figure}
\centering
\includegraphics[width=0.49\textwidth]{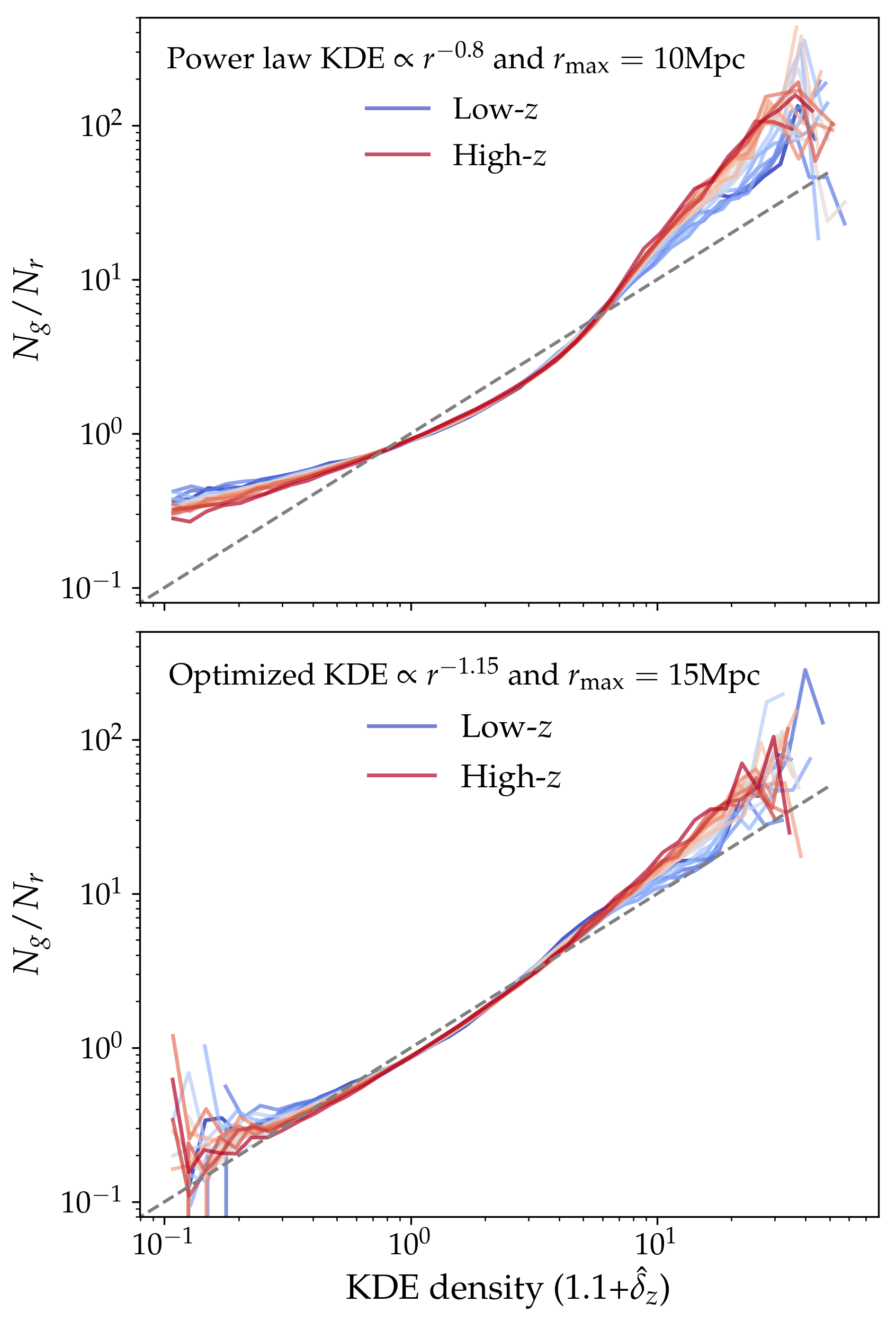}
	\caption{\textit{(Upper panel):} Ratio between the abundance of target galaxies and random points as a function of estimated KDE density, for a power law KDE $r\propto r^{-0.8}$ and $r_{\mathrm{max}}=10\mathrm{Mpc}$. The different redshift bins are color coded. If the KDE delivered a perfectly unbiased field estimate of the target galaxies, we would expect to find the dashed line relation. All galaxies have been used without tomographic bin selection to obtain a better estimate. The true redshift of all the target galaxies was used, while in a real data scenario one could only estimate this relation in the smaller calibration fields. \textit{(Lower panel):} Same as upper panel, but using an optimized KDE with $r_{\mathrm{max}}=15\mathrm{Mpc}$. The KDE is optimized from a function that combines a power law and an exponential truncation at small scales to deal with shot noise effects (see Fig. \ref{kde_optimize_formula}). The optimal parameters are found from a calibration field from  $\sim 3.5 \mathrm{deg}^{2}$ where redshifts for the target galaxies are known. It shows a more linear relation, although remains substantially nonlinear at the extremes of density.}
\label{mapping_powerlaw_10mpc}
\end{figure}

The optimization of the KDE works as follows. We write the probability of the optimized KDE parameters for redshift $z$ as
\begin{equation}
p(\alpha_z,r^{*}_z, \gamma_z|\theta, z) \propto p(\theta|z,\alpha_z,r^{*}_z, \gamma_z)~ p(\alpha_z,r^{*}_z, \gamma_z|z)
\end{equation}
where the last term is the prior on the parameters. Given a sample of targets with known redshifts from a calibration field, 
\begin{equation} 
\begin{split}
p(\theta|z,\alpha_z,r^{*}_z, \gamma_z) &\propto \prod_{i\in z} p(\theta_i|z,\alpha_z,r^{*}_z, \gamma_z)\\
&= \prod_{i\in z}\mathcal{B}\left(\hat{\delta}_z(\theta_i,\alpha_z,r^{*}_z, \gamma_z), \{b^{z}_k\}\right)
\end{split}
\end{equation}
where $p(\theta_i|z,\alpha_z,r^{*}_z, \gamma_z)$ is the probability of the $i$th galaxy at redshift $z$ from the calibration sample. We obtain this probability by biasing the KDE estimate $\hat{\delta}_z(\theta_i)$ using the $\mathcal{B}^{\rm true}_z$ from Equation (\ref{eq:mapping_noparam}), estimated using only the galaxies from the calibration sample. Note that since we know the true redshifts in the calibration sample we do not need to use the parametric form from Equation (\ref{eq:mapping_param_form}) but directly use the estimate from Equation (\ref{eq:mapping_noparam}).

When using a small patch of $\sim 3.5 \mathrm{deg}^{2}$ to optimize the kernel parameters, we take the average of the maximum-posterior parameters across all redshift bins as an estimate for the optimized KDE, since the constraining power in each redshift bin is weak. We use top hat priors $\alpha_z\in[-2,-0.5]$, $r^{*}_z\in[0.001,0.1]$ and $\gamma_z\in[-10,-2]$. The $\gamma$ parameter has little effect on the posterior so we fix it to its mean value of $\gamma=-4$ and run again. For a kernel limited to $r_{\mathrm{max}}=15\mathrm{Mpc}$ we find $<\alpha_z>=-1.15$ and $<r^{*}_z>=0.018\mathrm{Mpc}$. For a KDE limited to $r_{\mathrm{max}}=10\mathrm{Mpc}$ we find $<\alpha_z>=-1.0$ and $<r^{*}_z>=0.010\mathrm{Mpc}$. Note how a more aggressive power law is preferred when the size of the KDE is larger.
The lower panel of Figure \ref{mapping_powerlaw_10mpc} shows the biasing relation (estimated using all target galaxies in the simulation) for the optimized kernel with $r_{\mathrm{max}}=15\mathrm{Mpc}$, which is much closer to the ideal relation than the $\alpha=-0.8, r_{\mathrm{max}}=10\mathrm{Mpc}$ power-law kernel shown in the upper panel. This is both a consequence of having $2.25\times$ more area and of optimizing the kernel shape.  The bottom right panel of Fig. \ref{density_field_kde} shows the density field estimated with the optimized KDE with $r_{\mathrm{max}}=15\mathrm{Mpc}.$  This will be our default kernel for further testing.

We compute the FOM of \eqq{eq:fom} for several choices of kernel.  For this purpose (but not for the results in Section~\ref{sec:results}), we use the $\mathcal{B}^{\rm true}$  biasing function estimated using all galaxies. The median FOM value across redshift for the optimized kernel with  $r_{\mathrm{max}}=15\mathrm{Mpc}$ is 0.263, while for the power law we measure 0.240. For $r_{\mathrm{max}}=10\mathrm{Mpc}$ we find a median FOM of 0.266 for the optimized kernel and 0.219 for the power law. The average information gain per galaxy from optimizing the KDE is $2-5\%$, and we find the information to be similar for both $r_{\mathrm{max}}$ limits once the kernel shape has been optimized. We select as our default the optimized kernel with $r_{\mathrm{max}}=15\mathrm{Mpc}$ since it has a more linear and easier-to-model $\mathcal{B}$ shape. In general, the shape of the optimal KDE and the shape of the biasing relation (Fig. \ref{mapping_powerlaw_10mpc}) depend on the tracer sample density per unit comoving surface area, among other factors. Here we choose a tracer sample with constant comoving density, which minimizes the variation from this effect across redshift.  

\section{Sampling}
\label{sec:sampling}

Now we turn to the problem of sampling over the redshift and type probability distributions of populations of galaxies and their individual constituents, in the framework of the hierarchical Bayesian model described in the previous sections. It is complicated to simultaneously sample all variables from the joint posterior $p(\fractions, \redshifts, \types, \biases | \features, \positions)$ in \eqq{pzf3}. We will show, however, that it is feasible to draw samples from this posterior using a three-step Gibbs sampler, because the conditional posterior distributions of interest can be easily written and sampled. In SB19, the true values of the density field at each position were known, and hence there was no need to sample over the parameters $\{b_k\}$ defining the biasing function $\mathcal{B}\left(\hat{\delta}_z(\theta_i), \{b^{z}_k\}\right)$ relating the true galaxy density to the KDE (see Section \ref{sec:clustering}). 
This paper's implementation executes sampling over bias parameters, including the development of some key sampling strategies that will enable a future application to real data.    

We use information from all galaxies in the target sample to constrain the redshift and type probability distributions of the galaxy population. Additionally, the fully Bayesian nature of this scheme allows us to make use of prior information on the relevant quantities, when available. In this work, we will assume that we have access to a ``spectroscopic sample''  of the galaxies with known $z,t$, \eg\ from a complete spectroscopic survey of a random subsample of targets in a small region of the sky. We will also assume that we can identify a tracer population among the spectroscopic sample,
with the same selection as the corresponding tracers in the full sample. These subsamples will place an informative prior on the coefficients $\fractions$, and will also be important in sampling over the biasing function parameters described in Section~\ref{sec:clustering}.

\subsection{Three-step Gibbs sampler}
\label{sec:gibbs}
Each iteration of the Gibbs sampler comprises three steps which are (i) drawing a sample of $\fractions$ from $p(\fractions|\redshifts,\types,\biases,\features,\positions)$, (ii) drawing pairs of $z_i, t_i$ for each galaxy $i$ from $p(z_i,t_i|\fractions,\biases,F_i,\theta_i)$ using the newly drawn $\fractions$, and (iii) drawing a sample of the biasing function parameters $\biases$ for each redshift bin from 	$p(\biases_z|\fractions,\redshifts,\types,\features,\positions)$ given the $z_i$ assignments in step (ii). The conditional distributions can be derived from the joint distribution in \eqq{pzf3}. The first two steps of the sampler are as in SB19 and hence we skip the full derivation for brevity (see SB19 for more details), and the third step is new and is considered in more detail. 

\begin{enumerate}[label=(\roman*)]
	\item The conditional posterior on $\fractions$ depends on the counts of sources of $\redshifts$ and $\types$ (in the last iteration), $\counts = \{\nzt\}$ where $\nzt$ is the number of sources assigned to redshift $z$ and phenotype $t$, and it also depends on the prior information on $\fractions$, $p(\fractions)$:
		\begin{equation}
			p(\fractions|\redshifts,\types,\biases,\features,\positions) \propto p(\fractions) \prod_{z,t} f_{zt}^{\nzt} .
		\end{equation}
		The prior condition that $\sum f_{zt}=1$, and $0 \leq f_{zt} \leq 1$, allows us to write the conditional posterior on \fractions\ as a Dirichlet distribution. Following the derivation in SB19, if $\priorcounts=\{\mzt\}$ are the counts of the prior sample found at each $z,t$ pair, and we assume that each spectroscopic galaxy has been drawn independently from the distribution, then the prior distribution of $\fractions$ follows a Dirichlet distribution with parameters \priorcounts.  In this case the conditional posterior follows a Dirichlet on the data counts from the last iteration plus the prior counts:
\begin{equation}
p(\fractions|\counts) \sim \mathrm{Dir}(\counts + \priorcounts),
\end{equation}
\begin{align}
	\nonumber
	\mathrm{with \: \: Dir}(\counts) & \equiv (N + N_zN_t -1)!\, \delta_D\left(1-\sum_{zt}f_{zt}\right) \\
 & \phantom{=} \times  \prod_{z=1}^{N_z}\prod_{t=1}^{N_t}\frac{\Theta(f_{zt})f_{zt}^{n_{zt}}}{n_{zt}!}.
\nonumber \\
	\label{uninformative}
\end{align}
An important shortcoming of our scheme is that the spectroscopic sample will not usually satisfy the condition that all galaxy draws are independent, because it is taken from a limited sky area and thus subject to large-scale-structure variance.  The posterior will therefore not sample this form of variance. The addition of sample variance uncertainties into the prior sampling will be explored in a future publication. 

	\item For each galaxy, the posterior for the $z_i,t_i$ pair conditioned on \fractions and \biases  $~$is
		\begin{equation}
	p(z_i,t_i |\fractions,\biases,F_i,\theta_i) \propto \likeli_{it_i} \, f_{t_iz_i} \, \mathcal{B}\left(\hat\delta_{iz_i}(\theta_i), \biases_{z_i}\right)
		\end{equation}
		where apart from using the $\fractions$ obtained in the first step of the sampler (i), we make use of the measurement likelihood $\likeli_{it_i}$ and the clustering terms $\mathcal{B}$ discussed above. The sampling in this step (ii) will produce pairs of $z,t$ for each galaxy that constitute the next realization of $\counts = \{\nzt\}$, to be used in the step (i) of the next iteration of the Gibbs sampler.
	\item After we have $z$ assignments for all galaxies in the sample from step (ii), we can now separate galaxies into redshift bins according to those assignments. Then, for each redshift bin, the posterior on the biasing function of that bin conditioned on all other variables looks like: 
\begin{align} \label{biasing_sampling_equation}
	p(\biases_z|\fractions,\redshifts,\types,\features,\positions) & = p(\biases_z|\redshifts, \positions) \nonumber \\
	& \propto \prod_{i: z_i = z} \mathcal{B}\left(\hat\delta_{iz_i}(\theta_i), \biases_{z_i}\right). 
\end{align}
With the choice of parametric biasing function in \eqq{eq:biasing}, there is no direct sampling algorithm for this conditional posterior. We therefore use the following procedure: first, we a Metropolis-Hastings (MH)
Markov Chain Monte Carlo (MCMC) sampler for the conditional posterior in \eqq{biasing_sampling_equation} for each redshift bin where we restrict the galaxies to the spectroscopic sample.  Since the spectroscopic sample have fixed $z_i$, this chain can be run once, before the Gibbs sampling commences, and yields a sampling of the prior on bias parameters inferred from the spectroscopic sample (see appendix \ref{sec:app_a}).
		Next, at each iteration of the Gibbs sampler, we return the 5000th sample from an MH MCMC chain run on \eqq{biasing_sampling_equation} using all target galaxies currently assigned to a given redshift (we have performed this step with MCMC chains longer than 5000 steps, with consistent results).  The proposal distribution for this MH sampler is to draw at random from the output sampling of the prior.  
Effectively we are using the target sample for importance-sampling of the prior sample.  This procedure is a robust way to combine the prior and target conditionals without the need to tweak the proposal distributions or the parameter limits of the MCMC chains.      It is also very fast compared to step (ii) of the Gibbs sampler.
\end{enumerate}




\section{Results}
\label{sec:results}

We use the simulation described in \S\ref{sec:sims} to test the methodology developed throughout this work. The target sample for this Section is the third tomographic bin in Figure \ref{target_tracer_zdist}, which contains $\sim3.3\times10^6$ objects. The spectroscopic sample, for which redshift and type are assumed known, consists of all 11,000 target galaxies from one patch of sky with area $\sim3.5\mathrm{deg}^2$. These objects are used to estimate the prior probability $p(z,c)$ and obtain the sampled prior on the mapping function parameters $\mathcal{B}(\hat{\delta},\{b_i\})$  (see Section \ref{sec:sampling} for details about the sampling). 

The HBM method yields samples of the redshift and type posterior for each individual galaxy; the redshift and type posterior of the population; and the posterior of the biasing function parameters. We focus on the redshift population posterior, marginalizing over all other parameters, since this is what is usually needed in cosmological analyses of galaxy surveys. In particular, current and future weak lensing analyses are very sensitive to small biases in the mean redshift of the distribution, which can become the leading systematic uncertainty. Therefore, in analyzing our results, we define one quality metric to be the difference between the mean of each sample $j$ of our redshift posterior and the true mean from all the target galaxies,
\begin{equation}
\Delta z_{j} = \braket{z_{\mathrm{est,j}}} - \braket{z_{\mathrm{true}}}.
\end{equation}
Since we draw samples of the full redshift distribution posterior $f_z$, another useful metric that is sensitive to the distribution shape is the Kullback-Leibler divergence ($D_\mathrm{KL}$) between each sample and the true redshift distribution,
\begin{equation}
D_\mathrm{KL}(f_{z,j}^{\mathrm{est}}|| f_{z}^{\mathrm{true}}) = \sum_z f_{z,j}^{\mathrm{est}} \log\left(\frac{f_{z,j}^{\mathrm{est}}}{f_{z}^{\mathrm{true}}} \right).
\end{equation}
This is a measurement of the relative entropy between the true distribution and the recovered distribution, and can be used to see how much information the photometry and density estimates are adding with respect to the prior knowledge. A Kullback-Leibler divergence of 0 indicates that the two distributions in question are identical, and the lower its value the more similar the two distributions are. 

For each case we investigate, we sample $n(z)$ from three distributions: (1) the prior only; (2) the posterior from an HBM that only includes photometry information; and (3) the posterior from an HBM that includes both photometry and clustering information, marginalizing over the biasing parameters. We denote the HBM with photometry as $F$ (feature) and the HBM with photometry and clustering as $F+\delta$. The $F$ inference is essentially a rigorous application of the reweighting method of \citet{Lima2008}.

In the first part of this Section, we look into the impact of sample variance in the prior coming from the calibration sample. In the second part, we study how the method performs when the prior on the $p(z,t)$ probability from a calibration sample is modified and biased. For each case, we will show a violin plot of the posterior redshift distribution compared to the true distribution, the distribution of $\Delta z_{j}$ differences and the distribution of $D_\mathrm{KL}$ divergences.


\subsection{Sample variance in the prior}

As noted in Section~\ref{sec:gibbs}, we have adopted a Dirichlet prior on $p(z,c)$ that assumes that galaxies drawn from the small spectroscopic-sample patch of sky have independent phenotypes and redshifts.  This neglects sample variance from large-scale structure (hereafter just ``sample variance''), which adds noise to the estimated relative density of galaxies at given redshift and type $f_{tz}$ \citep{Cunha2012}. This effect is larger at lower redshifts, where the volume is smaller. 

Sample variance most importantly affects the density of types $p(t)$, where $p(z,t)=p(z|t)p(t)$, since the same phenotype would yield the same redshift regardless of where it is observed, provided the redshift distributions of phenotypes are narrow. However, we have seen in Fig. \ref{deep_som} that there are some phenotypes (deep cells) with wider redshift distributions, mostly due to color-redshift degeneracies. As a result, the redshift distribution $p(z|t)$ of these phenotypes is also affected by sample variance. The Dirichlet sampling of the prior, as presented in \S\ref{sec:sampling}, neglects sample variance uncertainty, but we expect the HBM method to reduce the effect of sample variance in the prior since the target population is much larger than the prior sample. Nevertheless, limited sampling or shot noise from the prior in any of the phenotypes can lead to a noise bias of $p(z|t)$, and make the HBM reconstruction imperfect. 

\begin{figure}
\centering
\includegraphics[width=0.45\textwidth]{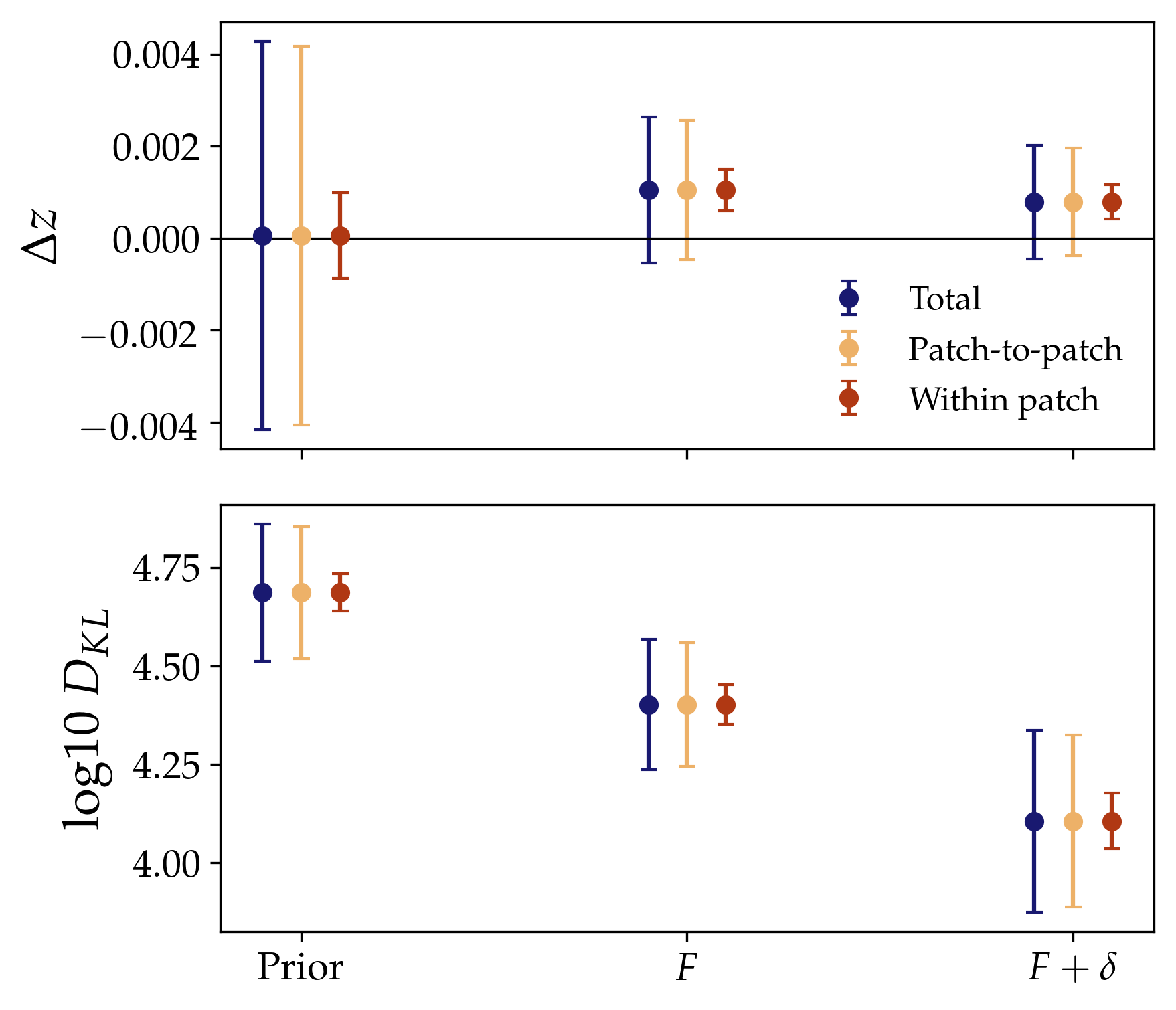}
\caption{Performance on the posterior redshift probability distribution for a hierarchical Bayesian model (HBM) with photometry and clustering information. Two metrics are shown, the bias in mean redshift distribution $\Delta z$ (upper panel) and the Kullback-Leibler divergence $D_\mathrm{KL}$ between the posterior samples and the true distribution (lower panel). The prior information comes from a small patch of $\sim3.5\mathrm{deg}^2$. We show results grouped in three blocks which show the results from drawing Dirichlet samples directly from the prior (labeled as Prior), from drawing samples using an HBM with only photometry ($F$) and from an HBM with both photometry and clustering ($F+\delta$). The total error budget (blue) is estimated from the standard deviation of samples drawn from HBM chains run in 11 randomly distributed patches of the same size. We also show the contribution to the total error of the sample variance (yellow, Patch-to-patch) and the mean internal variance of each chain (red, Within patch), finding the former one dominates the error budget in every case. The HBM reduces the sample variance uncertainty from the prior and significantly improves the recovered shape when also adding the clustering.}
\label{results_case1_sv0}
\end{figure}

To assess this sample variance, we randomly choose 11 calibration samples of $\sim3.5\mathrm{deg}^2$ each, and apply the HBM method to each, with and without using clustering information. In Figure \ref{results_case1_sv0}, we show the results of these runs in the two metrics defined above, \ie~the mean of the redshift distribution and the KL divergence compared to the truth. For each method of inference (prior-only, $F$, and $F+\delta$) we show the mean of both metrics over the 11 distinct spectroscopic patches, with three different uncertainty estimations: (1) the total standard deviation among all MCMC samples of all spectroscopic patches; (2) the standard deviation of the means of the 11 different prior patches; and (3) the standard deviation within the MCMC samples of one patch.  The Figure shows that:
\begin{itemize}
  \item The sample variance among patches (2) dominates the total uncertainty budget (1) in every case.
   \item The HBM ($F$) reduces the uncertainty in the estimation of the mean redshift, i.e. lessens the impact of spectroscopic sample variance, and also improves the $N(z)$ shape reconstruction (lower KL divergence values) compared to the prior-only inferences.
   \item The addition of the clustering further reduces the uncertainty in the mean redshift and improves the $N(z)$ reconstruction.
   \end{itemize}
 The HBM mean redshift uncertainty goes from $(0.0\pm4.2)\times10^{-3}$ in the prior to $(1.0\pm1.6)\times10^{-3}$ for HBM ($F$) and $(0.8\pm1.2)\times10^{-3}$ for HBM ($F+\delta$). The shape improves from a $\log_{10}\left(D_\mathrm{KL}\right)$ divergence of $4.69\pm0.17$ in the prior to $4.40\pm0.17$ and $4.11\pm0.23$ for HBM ($F$) and HBM ($F+\delta$), respectively.

\begin{figure}
\centering
\includegraphics[width=0.45\textwidth]{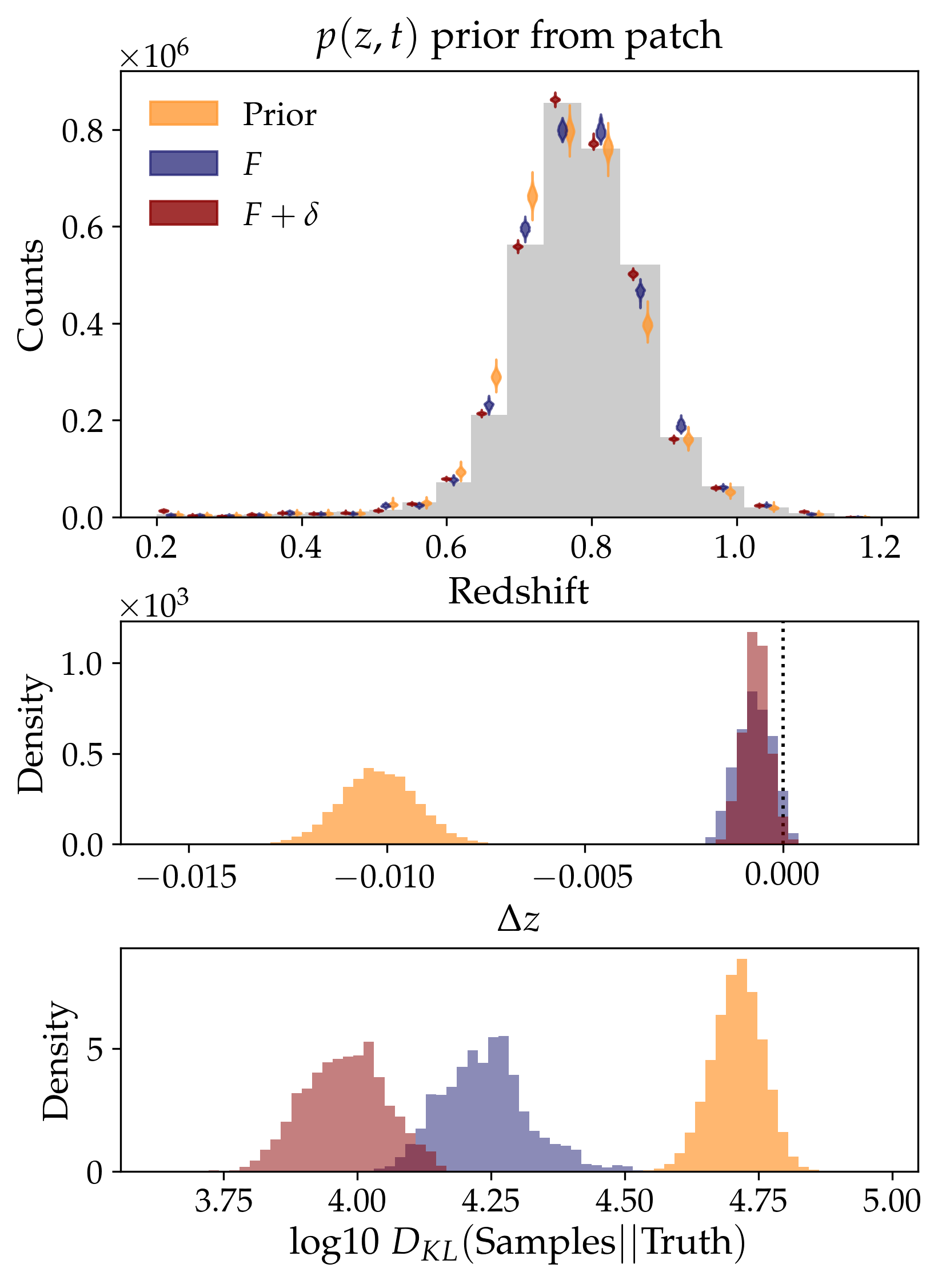}
\caption{Posterior redshift probability distribution, marginalized over type and when including clustering marginalizing over mapping function parameters. The prior is obtained from a small calibration patch with 10758 objects over an area of $\sim3.5\mathrm{deg}^2$. The three plotted distributions are obtained from: the prior; the posterior for an HBM with photometry only, $F$; and the posterior for an HBM with photometry and clustering $F+\delta.$. \textit{Top:} Shows violin plots for each distribution compared to the true redshift distribution (grey).  \textit{Middle:} Shows the posterior distribution of redshift bias  $\Delta z$ values. \textit{Bottom:} Shows the distribution of Kullback-Leibler divergence ($D_\mathrm{KL}$) between each sample and the true redshift distribution. The HBM ($F$) removes most of the redshift bias, since in this case the prior's redshift bias is primarily caused by biases in the type density $p(t)$ caused to the sample variance of the calibration patch. The addition of clustering sharpens the distribution and improves the overall shape, reducing the  $D_\mathrm{KL}$ divergence.}
\label{results_case1_sv}
\end{figure}

In Figure \ref{results_case1_sv} we randomly choose one of the spectroscopic-sample patches for the prior, and compare the posterior from running an HBM with photometry alone ($F$, blue), an HBM with photometry and clustering ($F+\delta$, red) and samples drawn from the Dirichlet prior on $p(z,t)$ (orange). The prior $p(z,t)$ has a mean redshift bias of $\Delta z=(-1.0\pm0.1)\times10^{-2},$ arising from LSS sample variance in this single sky patch. When running the HBM we find the bias reduced to $\Delta z=(-7.2\pm4.4)\times 10^{-4}$ with photometry alone and a bias of $\Delta z=(-6.7\pm3.2)\times 10^{-4}$ when adding clustering. In agreement with Figure \ref{results_case1_sv0}, we find the HBM with photometry alone, \ie\ reweighting \citep{Lima2008,Sanchez2014}, to be able to correct redshift biases that come from an LSS-biased type probability $p(t)$ (SB19). Since sample variance mostly changes $p(t)$,  having feature information is enough to remove most of the redshift bias. In this case that an unbiased spectroscopic sample is available for $\approx10^4$ galaxies, the addition of clustering information has little impact on the overall redshift bias.  Adding the clustering information does, however, further tighten the $\Delta z$ posterior distribution and also improves the shape of the redshift posterior, leading to a smaller $D_\mathrm{KL}$ divergence.  

\subsection{Biases in the prior} 

So far we have assumed our prior is an unbiased estimate of the underlying distribution in the spectroscopic patch, so it was only affected by sample variance. We now introduce several possible biases in the spectroscopic prior, mimicking some effects that we could find in real data, and analyze the ability of the HBM to overcome these biases. We will use same spectroscopic patch used in creating Fig. \ref{results_case1_sv}.

\subsubsection{Prior $p(z,t)$ with a redshift bias}
\label{pzt_bias}

\begin{figure}
\centering
\includegraphics[width=0.45\textwidth]{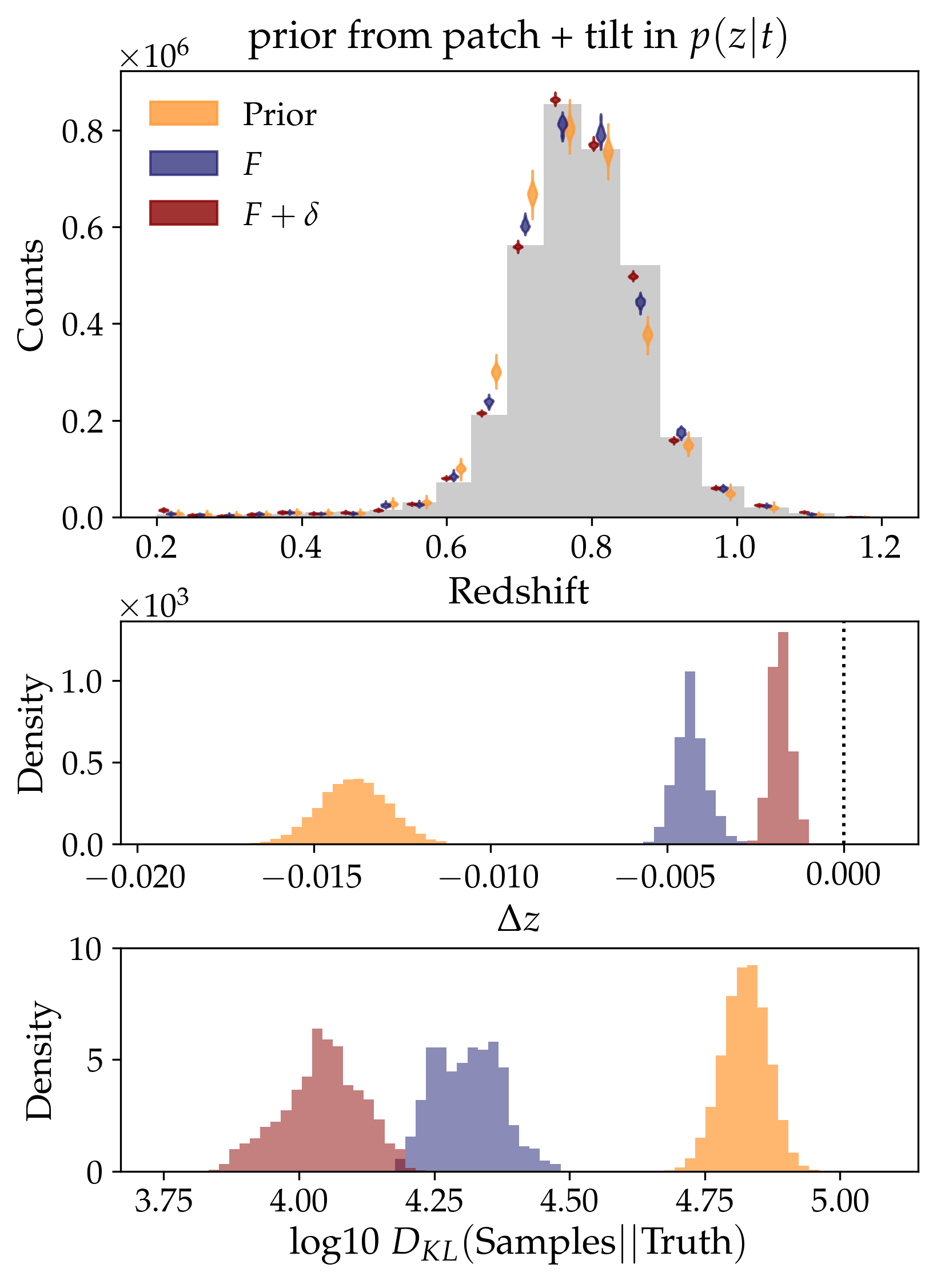}
\caption{Similar to Fig. \ref{results_case1_sv}. The prior, which is obtained from the same small calibration patch, is systematically biased in the conditional redshift probability of each type $p(z|t)$ towards low redshift as per \eqq{eq:linbias}. The HBM with photometry alone reduces the redshift bias by the same amount as in Fig. \ref{results_case1_sv}, since it only corrects redshift biases produced by a bias in $p(t)$. The remaining bias can only be corrected with the addition of clustering, which further reduces this bias and improves the redshift posterior shape.}
\label{results_case2_sv_biased}
\end{figure}

We add a systematic redshift bias for each phenotype/deep cell by altering its redshift distribution to
\begin{equation}
  p^\prime(z | t) \propto p(z | t) * (21-z), \quad z=1,2,\ldots,20.
\label{eq:linbias}
\end{equation}
Therefore, the prior $p(z,t)=p^\prime(z|t)p(t)$ now has a systematic bias towards low redshift.
Fig. \ref{results_case2_sv_biased} shows the HBM results for such a prior.  Drawing only from the prior, the mean redshift bias is  $\Delta z=(-1.4\pm0.1)\times10^{-2}$. The HBM with only photometry has a mean posterior redshift bias of $\Delta z=(-4.3\pm0.4)\times 10^{-3}$, while an HBM with photometry and clustering yields $\Delta z=(-1.8\pm0.3)\times 10^{-3}$. Note that the $F$-only HBM has corrected the same amount of redshift bias as in the previous case with unbiased prior ($\sim 0.01$ in $\Delta z$), \ie\ the sample variance, but cannot correct any of the systematic bias introduced in $p(z|t)$. The $F+\delta$ HBM, however can use the clustering information to further improve the $p(z|t)$ probability and reduce the total redshift bias. It also reduces the $D_\mathrm{KL}$ divergence, improving the overall shape.
 
\begin{figure}
\centering
\includegraphics[width=0.45\textwidth]{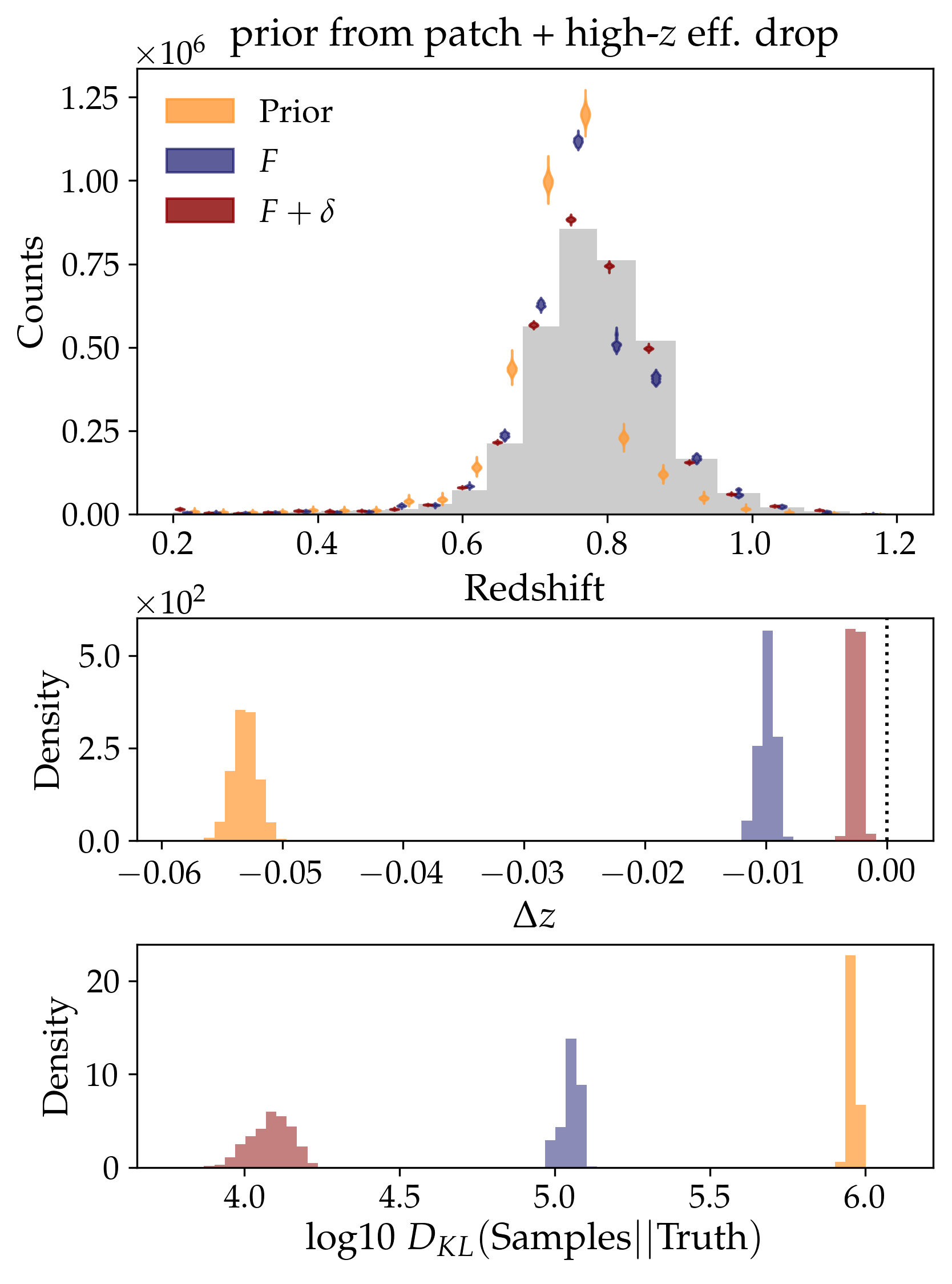}
\caption{Similar to Fig. \ref{results_case1_sv}. The prior mimics an hypothetical spectroscopic efficiency drop above redshift $z>0.8$ by reweighting the prior with a factor 0.2 in the 7 highest-redshift bins. The HBM with photometry is able to correct the redshift posterior in redshift bins far away from $z\sim0.8$, where the drop happens, by changing the density of deep cells whose redshift probability $p(z|t)$ does not cross $z\sim0.8$. It increasingly fails to correct the redshift distribution around $z\sim0.8$ since it cannot modify $p(z|t)$. Adding clustering significantly improves the redshift distribution, removing most of the redshift bias and largely improving the redshift distribution shape.}
\label{results_case3_sv_edrop}
\end{figure}

\subsubsection{Prior $p(z,t)$ with a redshift efficiency drop}

Spectroscopic surveys usually present sharp selection effects in redshift due to their limited wavelength coverage of the spectra. Using such survey to estimate the prior probability can bias the whole posterior redshift distribution of the weak lensing samples.  In this Section we use a prior $p(z,t)$ from a hypothetical spectroscopic survey with an efficiency drop above redshift $z>0.8$ (the 7 highest-redshift bins). We assume only $20\%$ of the galaxies in the last 7 redshift bins have been successfully measured with the failed measurement being simply discarded from the catalog, which we implement by multiplying by 0.2 the prior $p(z,t)$ in those bins.

Fig. \ref{results_case3_sv_edrop} shows that this efficiency drop creates a huge redshift bias in the prior of $\Delta z=(-5.3\pm0.1)\times10^{-2}$. For the $F$ HBM we find a redshift bias of $\Delta z=(-9.9\pm0.7)\times 10^{-3}$, while for the $F+\delta$ HBM we find $\Delta z=(-2.6\pm0.4)\times 10^{-3}$. The $F$ HBM is able to successfully correct redshift bins which are far away from where the efficiency drop happens ($z\sim0.8$) since there are many deep cells with a  very tight redshift-type relation. However, it has more difficulty recovering the redshift distribution closer to the drop, since it cannot update $p(z|t)$. Adding the clustering significantly improves the recovered shape, finding a much better $D_\mathrm{KL}$ divergence, and eliminates 95\% of the redshift bias from the prior.

\begin{figure}
\centering
\includegraphics[width=0.45\textwidth]{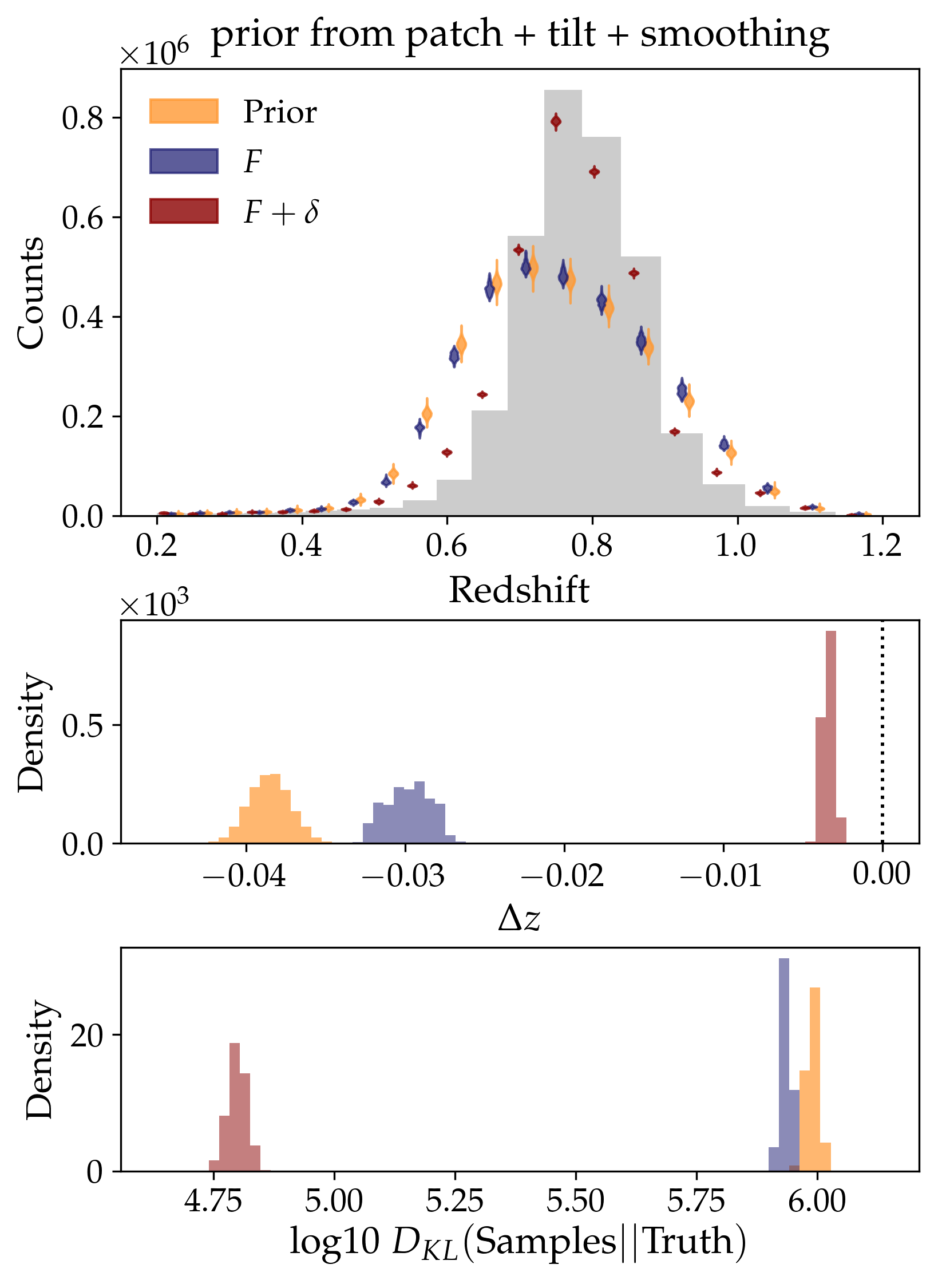}
\caption{Similar to Fig. \ref{results_case1_sv}. The prior is smoothed by convolving $p(z|t)$ with a top hat function of size 7 redshift bins, increasing the median redshift dispersion of the deep cells goes from $\sigma(z)=0.025$ to $\sigma(z)=0.1$, which reduces the correlation between type and redshift for all deep cells. In this case the HBM with photometry alone can barely modify the redshift distribution, since there is little correlation between type and redshift. In contrast, adding the clustering information remarkably improves the redshift distribution recovery and reduces most of the redshift bias. This shows that photometric redshift surveys with wider $p(z|t)$ estimation can be used instead of spectroscopic surveys when clustering is available.}
\label{results_case4_sv_smooth_biased}
\end{figure}

\subsubsection{Prior $p(z,t)$ with degraded $z-t$ correlation and biased}

So far we have assumed we have a calibration field with spectroscopic data that provide a tight redshift-color relation. Now we explore what happens if we loosen this assumption and pretend that the redshift information in the prior does not come from spectroscopy but from a hypothetical photometric redshift sample. This can be of interest in real data when spectroscopic redshifts can only sparsely populate the prior on $p(z,t).$ To mimic this effect, we convolve the conditional redshift probability for each type $p(z|t)$ with a top hat function with width of 7 redshift bins, which smooths the redshift probability. The median redshift dispersion of the deep cells goes from $\sigma(z)=0.025$ to $\sigma(z)=0.1$, significantly reducing the correlation between types and redshift. Furthermore, we add the same systematic redshift bias to each $p(z|t)$ as in Section~\ref{pzt_bias}. Note the sample variance in $p(t)$ is left unchanged.

Figure \ref{results_case4_sv_smooth_biased} shows the broadening effect in the prior, which now has a redshift bias of $\Delta z=(-3.9\pm0.1)\times10^{-2}$. The HBM with photometry alone, which can only modify the density of types, is barely able to change the redshift distribution since the correlation between redshift and type has been degraded, finding a redshift bias  of $\Delta z=(-3.0\pm0.1)\times10^{-2}$, and a very similar $D_\mathrm{KL}$ divergence. In contrast, adding the clustering remarkably improves the redshift bias and shape, leading to a very large decrease in both $D_\mathrm{KL}$ and $\Delta z$ metrics. In this case, we find a redshift bias of $\Delta z=(-3.4\pm0.3)\times 10^{-3}$. This result shows that, when clustering information is used in the HBM, photometric redshift estimates can be used instead of spectroscopic measurements, even if such photo-$z$ estimates are  imprecise and are systematically biased.

\section{Discussion}
\label{sec:discussion}

\begin{figure}
\centering
\includegraphics[width=0.45\textwidth]{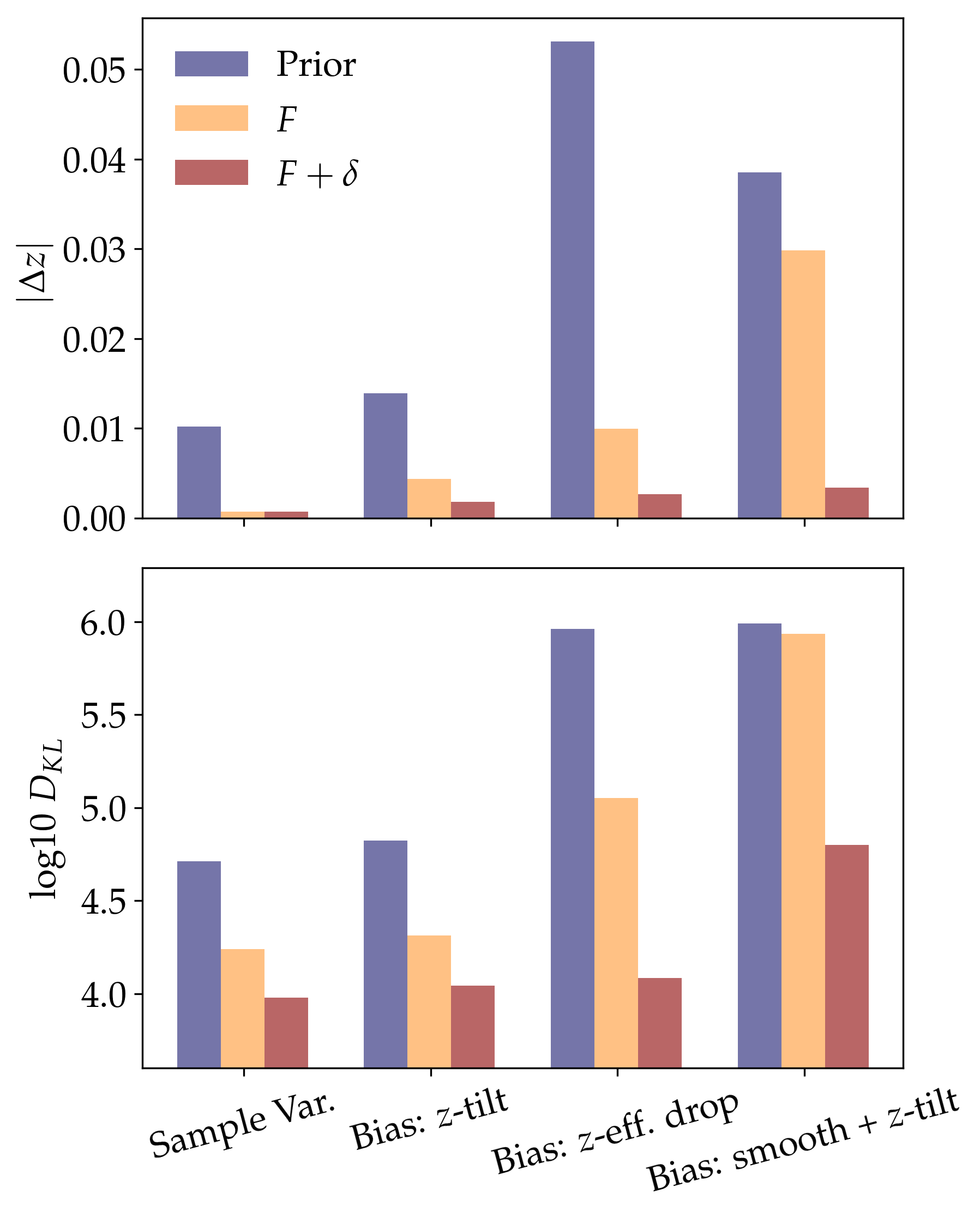}
\caption{A summary of the quality of $n(z)$ inferences obtained in this work. \textit{(Upper)}: the absolute redshift bias in the mean posterior redshift $|\Delta z|$. \textit{(Lower):} the Kullback-Leibler divergence $D_\mathrm{KL}$ between the posterior samples of $n(z)$ and the true distribution. We show the performance metrics are grouped in blocks of three, showing samples from: (1) the Dirichlet prior (labeled as ``Prior'') obtained directly from a spectroscopic survey of $\sim3.5\mathrm{deg}^2$; (2) from an HBM MCMC with only photometric information  (``$F$''); and (3) from an HBM with both photometry and clustering (``$F+\delta$''). The four cases studied are shown, one where the prior only has sample variance from the small patch, and three where the prior is further modified to introduce redshift biases. In all cases the HBM remarkably improves both the bias and shape of the posterior, and the best results are found with the addition of clustering information.}
\label{fig:summary}
\end{figure}

Figure \ref{fig:summary} presents a visual comparison of the two performance metrics ($\Delta z$ and $D_\mathrm{KL}$) obtained with three different inferences: (spectroscopic) prior from a small patch on the sky; the $F$ HBM with photometric information on the full sample; and the $F+\delta$ HBM including photometric information \emph{and} clustering against a tracer population. 
In the first case (``Sample Var.''~in the plot), where the prior has no biases but just sample variance, the $F$ and $F+\delta$ HBM methods show comparable results in terms of the mean redshift bias, but the clustering method performs better in recovering the shape of the redshift distribution (lower $D_\mathrm{KL}$ metric). In the other three cases, where biases are introduced in the prior, the HBM method with clustering always performs better in both metrics. Remarkably, for that method, the mean of the redshift distribution is always recovered with a precision of around $3\times10^{-3}$ or better, even when the redshift biases in the prior are larger than $5\times10^{-2}$. That is a very important result since accurate characterization of the mean of redshift is critical to cosmological analyses of weak gravitational lensing in imaging surveys.  Furthermore, the addition of clustering in the method always improves the reconstructed shape of the redshift distribution (lower $D_\mathrm{KL}$), which can also be very important for cosmology analyses: mischaracterization of the width or tails of a redshift distribution can be a source of systematic error for both weak lensing and galaxy clustering studies. 

Our results demonstrates the robustness of this method to several types of biases in the prior, chosen to mimic known shortcomings in real calibration samples. There is an ongoing discussion among the imaging surveys community about the reliability of different redshift samples and how biases in them are propagating into cosmological analyses and creating artificial tension with other cosmological probes  \citep{Troxel2018a,Joudaki2019,Asgari2019,Wright2019}. Some groups have relied on spectroscopic samples for their redshift calibration while others have used high-quality, many-band photometric redshifts instead. Spectroscopic samples provide accurate redshift information but can suffer from selection effects and efficiency problems, while high-quality photometric redshifts can have significant biases, especially at high redshift. We have demonstrated how the $F+\delta$ HBM method is robust to any of these effects, providing a rigorous way to propagate known priors into the posterior, as well as letting the clustering information overcome the priors and their potential inaccuracies.          

The success of this method in estimating redshift distributions to the accuracy needed for large cosmological surveys will still depend on the details of the survey. It is useful here to discuss how the application of this method to real data might differ from the simulations in this paper. First, in this work we have limited the redshift range of interest to be $0.2 < z < 1.2$, while in reality we will need to consider a larger redshift range \citep{Wright2019}. It could also be possible to consider an additional tracer sample at high redshifts, \eg\ a quasar sample. Second, the tracer sample used in this work is idealized in that it spans the entire redshift range of interest, and that we have assumed true redshifts for their galaxies. The latter assumption is not a problem as LRG samples have typical redshift errors of $\sigma_z \sim 0.02$ \citep{Rozo2015}, smaller than the redshift bin size chosen in our work ($\sim0.05$). Having a tracer sample not spanning the entire redshift range will reduce the constraining power of the method at the redshifts where we do not have tracers, but it will not result in any additional redshift bias, as demonstrated in SB19. Third,
the photometric noise likelihood function $p(F | t, \theta)$ has been determined comparing the truth to ``observed'' values in the full simulated target population, whereas in real data this function might be determined from injection simulations with a smaller number of realizations. Real data might therefore have weaker $F$-only reconstruction from added shot noise in $p(F|t,\theta),$ which would probably increase the degree of improvement that clustering information would yield.
Finally, one other difference is the area used in the application of the method. In this paper, we have used a sample of 1000~deg$^2$ of sky, while in the application of the method to data we can expect larger areas (\eg\ 5000 deg$^2$ in \des). A larger area overlap between the target and the tracer population will increase the constraining power of the method in the redshift range of overlap between these two samples, driving the biases of the $F+\delta$ HBM to lower levels than in our simulations.


Finally, we discuss the details of our implementation and the corresponding computational needs. For the run on the simulation, we define a total of 20 redshift bins equally spaced in comoving distance between $z\in[0.2,1.2]$, as well as 1024 phenotypes defined with an SOM from a $32\times32$ grid, so $f_{zt}$ has a total of 20,480 free parameters. Then, we have 100 free parameters in a biasing function with 5 parameters per redshift bin. And furthermore, for each target galaxy $i$, we have $z_i$ and $t_i$ as parameters, which amounts to $2\times3\times10^6$ free parameters. To save memory and improve speed,
we do not save the individual $z,t$ pairs for each target galaxy at every MCMC sample---the individual $(z_i,t_i)$ samples are aggregated into the number counts $N_{zt}$ necessary for the Gibbs sampling of $f_{zt}$. We parallelize the sampling of the individual $z,t$ of each galaxy in 334 chunks defined by healpy pixels of nside=$2^5$, and we parallelize the MCMC chain for the biasing parameters by assigning each redshift bin to its own thread. On average, a full iteration of the chain which samples all parameters using the 3 Gibbs intermediate steps takes 9 seconds using 334 parallel jobs, which gives about 400 iterations per hour. The method can be parellelized further for more speed, as that step is the limiting factor. Overall, the Gibbs sampling scheme is simple but has the drawback of yielding long correlation lengths, so that more iterations are needed to get a given number of independent samples. A Hamiltonian Monte Carlo (HMC) implementation is possible, and would yield practically independent samples which would result in a speed up of the method. This HMC implementation may be needed to make the method scalable for next generation surveys such as LSST.

\section{Summary and conclusions}
\label{sec:conclusions}

SB19 presented a hierarchical Bayesian model which can naturally combine the three main sources of information for estimating the redshift probability distributions of galaxies and samples of galaxies in a wide-field survey. These three main sources of information are: prior information, which comes from a subset of galaxies with well measured photometric and (typically) spectroscopic properties; broad-band photometry for the galaxies in the wide-field sample; and the clustering of such galaxies against a tracer population with precise and accurate redshift estimates. All these sources of information have been used separately in the past, but this is the first method to combine them in a unified and consistent way. In SB19, the main features and potential advantages of the method were demonstrated on a simple set of simulations, but the actual capabilities of it were not assessed, as they depend upon some important pieces that are needed for its application to real data, like realistic clustering properties and the marginalization over biasing functions in the usage of that clustering information.        

In this work, we have expanded the HBM approach of SB19 to include the additional methods needed for its application to the analysis of galaxy survey data. The HBM assumes that the galaxies come from a Poisson sampling of an underlying density field; in this work, we characterize this field as a kernel density estimator $\hat\delta(\theta)$ applied to a tracer galaxy population with known redshifts, then modified by some parametric biasing function $\mathcal{B}(\hat\delta, b).$
We have detailed here how such a biasing function can be constructed, with appropriate freedom to vary with redshift, and how we can sample and marginalize over it using prior information from spectroscopic information over a limited area of the sky. 

Moving beyond the simplistic simulations in SB19, we have now tested the methodology on the public MICE2 simulation, a mock galaxy catalog created from a lightcone of a dark-matter-only N-body simulation with $\approx200$ million galaxies over an octant of the sky. This simulation features realistic galaxy clustering and galaxy properties, and this allows us to work in a scheme where we can fully employ the phenotype approach proposed in SB19. Under that approach, we assume we have a sample with deep photometry and extra bands to define galaxy phenotypes, and a wide sample with noisier photometry and only a subset of optical bands as observations. We use two self-organizing maps (SOMs) to characterize the properties of these samples, and we use galaxies with best matching cells in both SOMs to accurately calibrate the likelihood probability that relates wide-field observations and phenotypes, as we would do in real data. 

In applying the method to a tomographic bin defined in the simulation, we always assume there is a small region of the sky (of about 3~deg$^2$) for which the galaxy properties, phenotype and redshift, are well known. We use this set of galaxies as a prior, both for the phenotype and redshift probability distribution and for the biasing function needed for the addition of clustering information from a tracer population. With this setup, we apply the methodology under different cases, comparing the results obtained with and without clustering information in the method and those from just the prior information. As metrics, we use the difference in the mean of the derived and true redshift distributions for the sample, which is arguably the most important quantity for weak lensing analyses, as well as the Kullback-Leibler divergence, which measures the differences in the shapes of the true and recovered redshift distributions.       

When the prior comes with perfect knowledge of a small patch of sky, \ie unbiased but with sample variance, the HBM method both with and without clustering information perform similarly well in terms of the mean redshift of the population. This is expected, also consistent with SB19, as sample variance mostly changes the phenotype distribution, and that can be recovered in the HBM without the need of clustering information. The shape of the redshift distribution is, however, better recovered when using clustering information.

Clustering information is shown to be very powerful when the redshift information from the small area is biased or incomplete, as is happening in real spectroscopic samples.  In such tests, the addition of clustering to the HBM improves both the mean and the shape of redshift distributions.  We have demonstrated this with simulations of a gentle coherent bias in the redshift assignments, in the case of uncompensated high-redshift incompleteness of spectroscopy, and in a case with spuriously broad spectroscopic assignments (as one might expect from photometric reference samples).  In these cases the HBM with clustering reduced the bias in the sample's mean redshift by a factor of 2--10 compared to photometry-only constraints.  The error in the full redshift distribution $n(z)$ is reduced by factors of 3--20, as measured by the Kullback-Liebler divergence.

One shortcoming of the current implementation of the HBM is that we do not account for correlations between the redshifts and phenotypes of the spectroscopic sample induced by large-scale structures in the spectroscopic sampling patch. With the current methods, this variance will need to be estimated with simulations such as the one done here. However, we are currently working on a way to add sample variance uncertainties into the prior sampling. In addition, future renditions of the HBM could be able to treat the density fluctuation field as a stochastic variable and hence include the LSS correlations.

The tests performed in this work provide demonstration that the method depicted in SB19, with the generalizations presented here, can be used in realistic conditions, and it can still be very powerful at resolving biases that are potentially present in prior samples, even after marginalizing over biasing functions in the addition of clustering information. The method does not guarantee an unbiased posterior, but it uses all the information at hand to reduce prior biases, and even in all tests performed here, some of which are extreme cases of biased priors, the final biases in the posterior are of order $10^{-3}$ in the mean of redshift distributions.  Obtaining a trustworthy $n(z)$ estimation of this accuracy in real survey data would be a milestone for the control of redshift systematic uncertainties in future weak lensing and galaxy clustering analyses.

\section*{Acknowledgements}

The authors thank Boris Leistedt, Daniel Gruen, Justin Myles and Alexandra Amon for helpful conversations about this topic. AA and EG were supported by MINECO grants CSD2007-00060 and AYA2015-71825, LACEGAL Marie Sklodowska-Curie grant No 734374 with ERDF funds from the European Union Horizon 2020 Programme. CS and GMB were supported by grants AST-1615555 from the US National
Science Foundation, and DE-SC0007901 from the US Department of Energy. IEEC is partially funded by the CERCA program of the Generalitat de Catalunya.
The MICE simulations have been developed at the MareNostrum supercomputer (BSC-CNS) thanks to grants AECT-2006-2-0011 through AECT-2015-1-0013. Data products have been stored at the Port d'Informacio Cientifica (PIC), and distributed through the CosmoHub webportal (cosmohub.pic.es).



\bibliographystyle{mnras}
\bibliography{library}


\appendix

\section{Prior and posterior of KDE biasing functions}
\label{sec:app_a}

\begin{figure*}
\centering
	\includegraphics[width=0.7\textwidth]{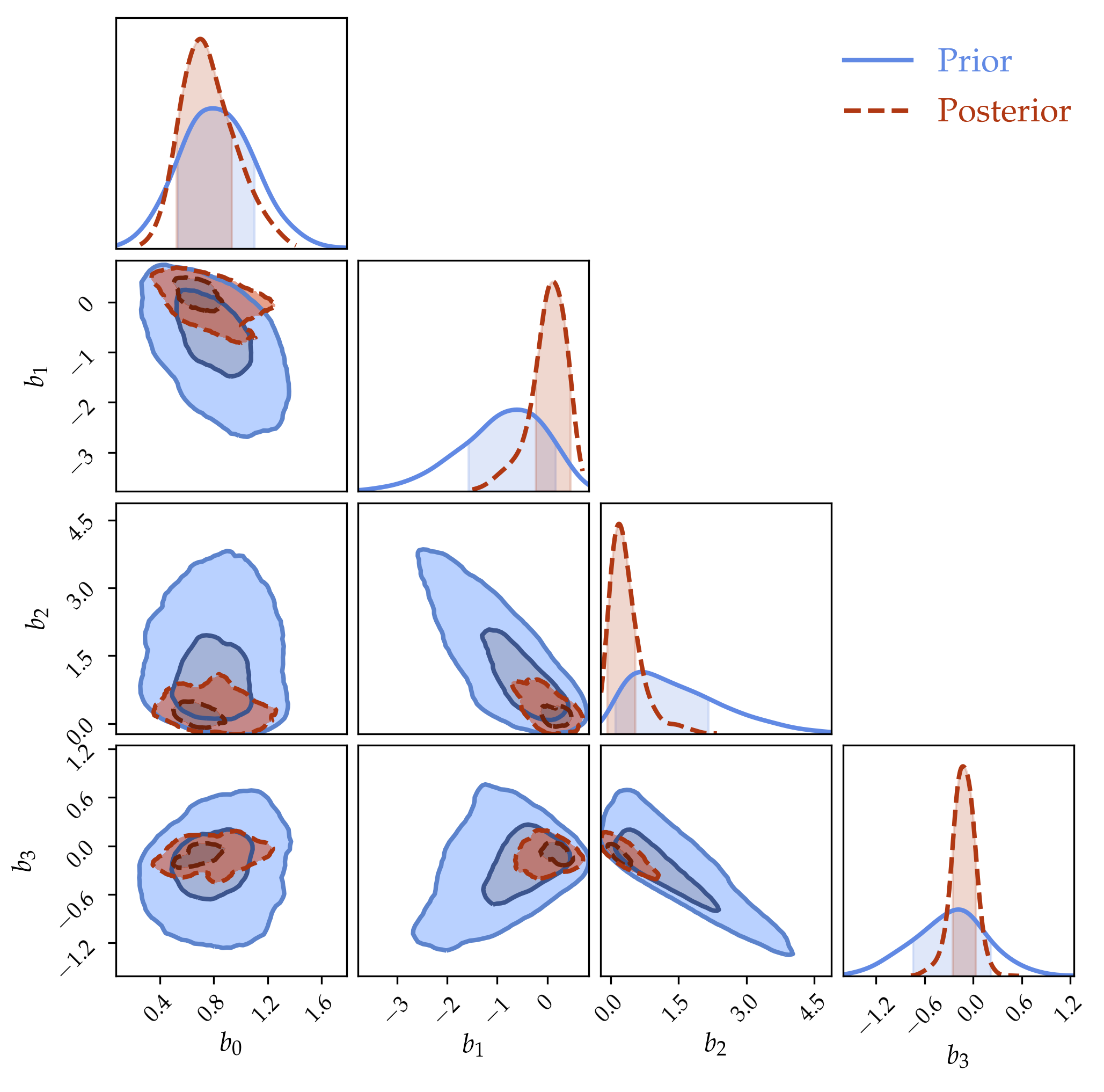}
	\caption{Prior and posterior of the biasing functions, parametrized as in Equation (\ref{eq:biasing}), in one random redshift bin (bin 4). The posterior appears to be tighter than the prior, showing how the HBM method uses information from the entire sample to characterize these mapping functions. }
\label{fig:app_a}
\end{figure*}

In this work, we use a galaxy tracer population to estimate the density field from which target galaxies are drawn from, using a kernel density estimation (\S \ref{sec:clustering}). However, as tracer and target populations can be different, and because of effects such as shot noise in the tracer population, we need a mapping function that relates the field estimated from tracers and the field from which target galaxies have been drawn from. As outlined in Section \ref{sec:sampling}, the biasing functions need to be sampled and marginalized over in the Gibbs process of the HBM. For that sampling, we use information from a small set of galaxies with true redshift information as a prior for the Gibbs sampling. In this work, in order to avoid being limited by sample variance in the estimation of this prior for biasing functions, we assume such functions have a smooth redshift dependence and we join 4 redshift bins from that prior sample at the time of running the corresponding MCMC chains. Then, we effectively use the same prior for 4 adjacent redshift bins in the Gibbs sampling process. Other than reducing sample variance, this procedure also makes the prior more robust to biases in the redshift estimation of the galaxies used in the prior. Figure \ref{fig:app_a} shows an example of the prior and posterior of such biasing functions, parametrized as in Equation (\ref{eq:biasing}), in one random redshift bin. One can see the posterior given by the HBM chain to be much tighter than the prior, showing how the HBM method is self-calibrating the biasing functions from the wide data in the simulation. The figure here shows one example redshift bin, but this is generally true for all bins considered in this work.


\bsp	
\label{lastpage}
\end{document}